\documentclass[aps,prb,reprint,superscriptaddress]{revtex4-1}
\usepackage{amsmath}
\usepackage{multirow}
\usepackage{graphicx}
\usepackage{epstopdf}
\usepackage{tabularx, booktabs}
\usepackage{pbox}
\newcolumntype{Y}{>{\centering\arraybackslash}X}
\usepackage[usenames,dvipsnames]{color}
\usepackage[caption=false,position=b,singlelinecheck=off,font=normalsize,labelfont=bf,justification=justified]{subfig}
\usepackage{hyperref}

\hypersetup{
	colorlinks=true,
	linkcolor=blue,
	filecolor=black,      
	urlcolor=blue,
	citecolor=blue,
}
\graphicspath{ {Figures_PDF/} }

\newcommand{\Li}{$\beta$-Li$_{2}$IrO$_3$}

\begin{document}

\title[]{Pressure-induced structural dimerization in the hyperhoneycomb iridate $\mathbf{\beta}$-Li$_2$IrO$_3$ at low temperatures} 

\author{L.S.I. Veiga}
\email[]{l.veiga@ucl.ac.uk}
\affiliation{London Centre for Nanotechnology and Department of Physics and Astronomy, University College London, Gower Street, London WC1E6BT, UK}

\author{K. Glazyrin}
\affiliation{Deutsches Elektronen-Synchrotron (DESY), Hamburg 22607, Germany}

\author{G. Fabbris}
\affiliation{Advanced Photon Source, Argonne National Laboratory, Argonne, Illinois 60439, USA}

\author{C.D. Dashwood}
\affiliation{London Centre for Nanotechnology and Department of Physics and Astronomy, University College London, Gower Street, London WC1E6BT, UK}

\author{J.G. Vale}
\affiliation{London Centre for Nanotechnology and Department of Physics and Astronomy, University College London, Gower Street, London WC1E6BT, UK}

\author{H. Park}
\affiliation{Physics Department, University of Illinois at Chicago, Chicago, IL 60607}
\affiliation{Materials Science Division, Argonne National Laboratory, Lemont, Illinois, 60439, USA}
 
\author{M. Etter}
\affiliation{Deutsches Elektronen-Synchrotron (DESY), Hamburg 22607, Germany}

\author{T. Irifune}
\affiliation{Geodynamics Research Center, Ehime University, Matsuyama, 790-8577, Japan}
\affiliation{Earth-Life Science Institute, Tokyo Institute of Technology, Tokyo, Japan}

\author{S. Pascarelli}
\affiliation{European Synchrotron Radiation Facility, 71 Avenue des Martyrs, 38043 Grenoble, France}

\author{D.F. McMorrow}
\affiliation{London Centre for Nanotechnology and Department of Physics and Astronomy, University College London, Gower Street, London WC1E6BT, UK}

\author{T. Takayama}
\affiliation{Max Planck Institute for Solid State Research, Heisenbergstrasse 1, 70569 Stuttgart, Germany}
\affiliation{Department of Physics and Department of Advanced Materials, University of Tokyo, 7-3-1 Hongo, Tokyo, 113-0033, Japan }

\author{H. Takagi}
\affiliation{Max Planck Institute for Solid State Research, Heisenbergstrasse 1, 70569 Stuttgart, Germany}
\affiliation{Department of Physics and Department of Advanced Materials, University of Tokyo, 7-3-1 Hongo, Tokyo, 113-0033, Japan }

\author{D. Haskel}
\email{haskel@aps.anl.gov}
\affiliation{Advanced Photon Source, Argonne National Laboratory, Argonne, Illinois 60439, USA}

\date{\today}

\begin{abstract}

A pressure-induced collapse of magnetic ordering in {\Li} at $P_m\sim1.5- 2$ GPa has previously been interpreted as evidence for possible emergence of spin liquid states in this hyperhoneycomb iridate, raising prospects for experimental realizations of the Kitaev model. Based on structural data obtained at \emph{room temperature}, this magnetic transition is believed to originate in small lattice perturbations that preserve crystal symmetry, and related changes in bond-directional anisotropic exchange interactions. Here we report on the evolution of the crystal structure of {\Li} under pressure at low temperatures ($T\leq50$ K) and show that the suppression of magnetism coincides with a change in lattice symmetry involving Ir-Ir dimerization. The critical pressure for dimerization shifts from 4.4(2) GPa at room temperature to $\sim1.5-2$ GPa below 50 K. While a direct $Fddd \rightarrow C2/c$ transition is observed at room temperature, the low temperature transitions involve new as well as coexisting dimerized phases. Further investigation of the Ir ($L_3$/$L_2$) isotropic branching ratio in x-ray absorption spectra indicates that the previously reported departure of the electronic ground state from a $J_{\rm{eff}}=1/2$ state is closely related to the onset of dimerized phases. In essence, our results suggest that the predominant mechanism driving the collapse of magnetism in {\Li} is the pressure-induced formation of Ir$_2$ dimers in the hyperhoneycomb network. The results further confirm the instability of the $J_{\rm{eff}}=1/2$ moments and related non-collinear spiral magnetic ordering against formation of dimers in the low-temperature phase of compressed {\Li}.

\end{abstract}

\maketitle

\section{Introduction}

Honeycomb-based systems with strong spin-orbit (SO) coupling have taken central stage in condensed matter physics due to their potential to realize a Kitaev quantum spin liquid (QSL) state~\cite{Kitaev06, Khaliullin09, Kim12, Kimchi14}. This non-trivial ground state emerges from the presence of bond-directional exchange anisotropy rooted in the strong SO interaction at the heavy transition-metal ion sites and geometrical frustration, inherent to the honeycomb lattice structure. The conflicting nature of the bond-dependent anisotropy leads to strong magnetic frustration, preventing occurrence of long-range magnetic ordering and favoring long-range quantum entanglement between the effective local moments~\cite{Kitaev06, Balents2010}. The resulting Kitaev QSL state is topologically protected, supports fractional non-Abelian anyons excitations and its realization holds promise for applications in quantum information and quantum computation areas~\cite{Kitaev06}.

Among the material candidates for possible realization of Kitaev physics, the honeycomb-based iridates (Ir$^{4+}$, $5d^5$) and ruthenates (Ru$^{3+}$, $4d^5$) are of particular interest due to their close proximity to bond-dependent ferromagnetic Ising interaction between their $J_{\rm{eff}}=1/2$ moments~\cite{Rau15, Winter16, Winter2017}. In this context, extensive experimental and theoretical studies have been devoted to the 2-dimensional (2D) honeycomb structures, such as $\alpha$-Li$_2$IrO$_3$~\cite{Singh12, Knolle14}, Na$_2$IrO$_3$~\cite{Singh10, Singh12, Choi12}, and $\alpha$-RuCl$_3$~\cite{Plumb2014, Johnson2015, Banerjee16}. However, deviations from ideal edge-sharing octahedral geometry mean that their ground state is not a QSL and instead, long-range magnetic order is stabilized at low temperatures and zero external magnetic field~\cite{Liu2011, Ye12, Williams16, Johnson2015, Cao2016, Banerjee16}. A further breath in the search for candidates of the Kitaev model emerged with the discovery of the 3-dimensional (3D) analogs of the 2D-honeycomb $\alpha$-Li$_2$IrO$_3$: {\Li}~\cite{Takayama15} and $\gamma$-Li$_2$IrO$_3$~\cite{Modic14}. Although these systems also order magnetically at low temperatures~\cite{Biffin14, Biffin14PRL}, it has been suggested that these materials are located in closer vicinity to the spin-liquid regime~\cite{HSKim15, Katukuri2016}, triggering enormous experimental efforts to test their properties under internal/external stimuli.

Within this context, evidence in favor of spin-liquid states under applied pressure, external magnetic field or chemical substitution has been put forward for $\alpha$-, $\beta$-, $\gamma$-Li$_2$IrO$_3$ and $\alpha$-RuCl$_3$. For example, a collapse of magnetic ordering and the subsequent emergence of a QSL-like phase has been reported for $\alpha$-RuCl$_3$ under magnetic fields applied parallel to the honeycomb planes~\cite{Johnson2015, Sears2017, Hentrich2018, Zheng2017}. Similar field-induced suppression of incommensurate spiral magnetic order has been observed for both 3D polytypes of Li$_2$IrO$_3$~\cite{Ruiz2017, Modic2017}. Very recently, the observation of a QSL has been reported in a hydrogen-intercalated iridate H$_3$LiIr$_2$O$_6$~\cite{Kitagawa2018}, with an apparent QSL phase replacing the complex magnetic order of the parent compound $\alpha$-Li$_2$IrO$_3$.    

External pressure constitutes another extensively employed tool to tune the magnetic ground state of these materials. Indeed, the apparent collapse of magnetic ordering in {\Li}~\cite{Takayama15, Veiga2017} ($T_N=38$ K) and $\gamma$-Li$_2$IrO$_3$~\cite{Breznay17} ($T_N=38$ K) at unexpectedly low pressures ($P_m\sim1.5-2$ GPa) has generated significant interest in understanding what kind of local moment interactions develop in the high-pressure phase, what mechanism drives them, and whether such interactions would ultimately lead to a QSL state. In {\Li}, X-ray magnetic circular dichroism (XMCD) reveals that the ferromagnetic response in an applied magnetic field disappears at $\sim2$ GPa, the same pressure range where X-ray absorption spectroscopy (XAS) shows a reconstruction of the electronic structure and departure from the strong SO coupling limit~\cite{Veiga2017}. Muon spin relaxation ($\mu$SR) measurements also reveal a breakdown of magnetic ordering at similar pressures ($\sim1.4$ GPa) and emergence of a new ground state marked by the coexistence of disordered states, namely, spin liquid and spin glass~\cite{Majumder2018}. In $\gamma$-Li$_2$IrO$_3$, single crystal x-ray diffraction (XRD) measurements of the $a$-axis lattice parameter under pressure show a smooth evolution across the low temperature magnetic transition, implying that structural distortions were not at play~\cite{Breznay17}. However, studies of the first-order, pressure-induced structural phase transition in {\Li} at $P_S\sim 4.4(2)$ GPa at \emph{room temperature} show a strong non-linear response of the $b$-axis compressibility across the structural boundary while the compressibility of the $a$-axis remains rather linear~\cite{Veiga2017}. The high-pressure monoclinic structure features significantly shortened Ir-Ir bond length within the zigzag chain, and the anticipated formation of Ir$_2$ dimers at $\sim4$ GPa was further confirmed by recent neutron diffraction experiments at room temperature~\cite{Takayama2019}. Note that the pressure-induced collapse of magnetism and changes in electronic structure were all observed in experiments conducted below 10 K. The critical pressure of the structural transition at low temperatures remains to be clarified.


Here, we address this gap of knowledge by revealing that {\Li} undergoes a dimerization transition at $P_S=1.5$ GPa at $T\leq50$ K, significantly below the critical pressure of $P_S\sim4.4(2)$ GPa observed at room temperature. While a direct $Fddd \rightarrow C2/c$ transition is observed at room temperature, the low temperature transitions involve new as well as coexisting dimerized phases, where Ir$_2$ dimers are formed in either of the $X, Y$ and $Z$-bond directions of the hyperhoneycomb structure. X-ray absorption near edge structure (XANES) measurements at Ir $L$-edges reveal that the pressure-induced suppression of the isotropic Ir ($L_3/L_2$) branching ratio and breakdown of $J_{\rm{eff}}=1/2$ state are closely related to the formation of the dimerized phases at high pressures. Our results, therefore, confirm that formation of Ir$_2$ dimers, possibly involving spin-singlet states, is the leading mechanism driving the collapse of magnetism in pressurized {\Li}. Formation of spin-singlet dimers is consistent with the very small paramagnetic response in applied field ($T=5$ K and $\mu_0H=4$ T) observed in our XMCD studies of compressed {\Li}~\cite{Veiga2017}.  



\section{Experimental Methods}\label{sec:experiment}

Single crystals and polycrystalline samples of {\Li} were synthesized by solid state reaction as described in Ref.~\onlinecite{Takayama15}. High pressure single crystal XRD experiments were performed at P02.2 beam line of PETRA III/DESY. A symmetric diamond anvil cell (DAC-Mao type) was used; this included Boehler-Almax diamonds (400 $\mu$m culet size) along with a rhenium gasket preindented to 80 $\mu$m. Single crystals with different dimensions ($\sim 10-20$ $\mu$m of thickness) were loaded in a 200 $\mu$m hole in the Re gasket, together with ruby spheres for \emph{in situ} pressure calibration. Neon gas was used as the pressure-transmitting medium, and the single crystal XRD patterns were collected with a Perkin Elmer XRD1621 detector. The low temperature measurements were performed using a He cold finger cryostat. The x-ray energy was tuned to 42.8 keV ($\lambda=0.2894$ {\AA}). The XRD patterns were analyzed using Crysalis Pro~\cite{Crysalispro}, Olex2~\cite{Olex2} and Jana2006~\cite{Jana2006} softwares. The quality of the crystals was tested prior to each individual loading. We found that most of the crystals had a small twin (intrinsic feature of the material); however, by means of the micro-diffraction measurements major domains could be selected and their structure solved. Information on the lattice parameters, atomic positions and other details of the structural solution are present in the Supplemental Material~\footnote{See Supplemental Material at [URL will be inserted by publisher] for additional information on the structural solution.}.


XANES measurements under pressure were performed at the undulator beamline 4-ID-D of the Advanced Photon Source, Argonne National Laboratory. X-rays were tuned to the Ir $L_{2,3}$ edges ($E=12.824$ keV and $11.215$ keV, respectively). A diamond anvil cell fitted with compression and decompression membranes was prepared with nanopolycrystalline diamonds (NPD) to mitigate the distortion of XANES data by diamond Bragg peaks~\cite{Irifune2003, Ishimatsu2012}. The decompression membrane was used to prevent pressure increases on cooling. Culet size of 400 $\mu$m was used. The DAC was mounted on a variable-temperature insert of a superconducting magnet for room and low-temperature measurements ($300$ K and $50$ K). Powders of {\Li} were loaded into a 200 $\mu$m hole in a stainless steel gasket pre-indented to 52 $\mu$m, together with ruby spheres for \emph{in situ} pressure calibration and mineral oil as the pressure-transmitting medium.

\section{Results}
\subsection{Pressure-temperature (P-T) dependence of the crystal structure}

Single crystal x-ray diffraction measurements were performed under variable P-T conditions to shed light on the mechanism behind the pressure-induced electronic and magnetic transitions observed in previous studies~\cite{Veiga2017, Majumder2018}. At room temperature, {\Li} undergoes a structural phase transition from the orthorhombic $Fddd$ space group to the monoclinic $C2/c$ structure at 4.4(2) GPa~\cite{Veiga2017}. The low-pressure, orthorhombic structure is characterized by an anisotropic compression of its lattice parameters as a function of pressure, with a faster contraction of the $b$-axis as a result of the rapid compression of the $X, Y$ bonds that form the zig-zag chains. On the other hand, the high-pressure monoclinic phase, which maintains the same hyperhoneycomb network of edge-shared IrO$_6$ octahedra, is described by a modulation of the Ir-Ir bond lengths. In this case, the $X$- and $Y$-bonds are no longer equivalent and the latter is dramatically shortened, indicating formation of Ir-Ir dimers in the hyperhoneycomb network.
\begin{figure}
	\centering
	\includegraphics[width=\linewidth]{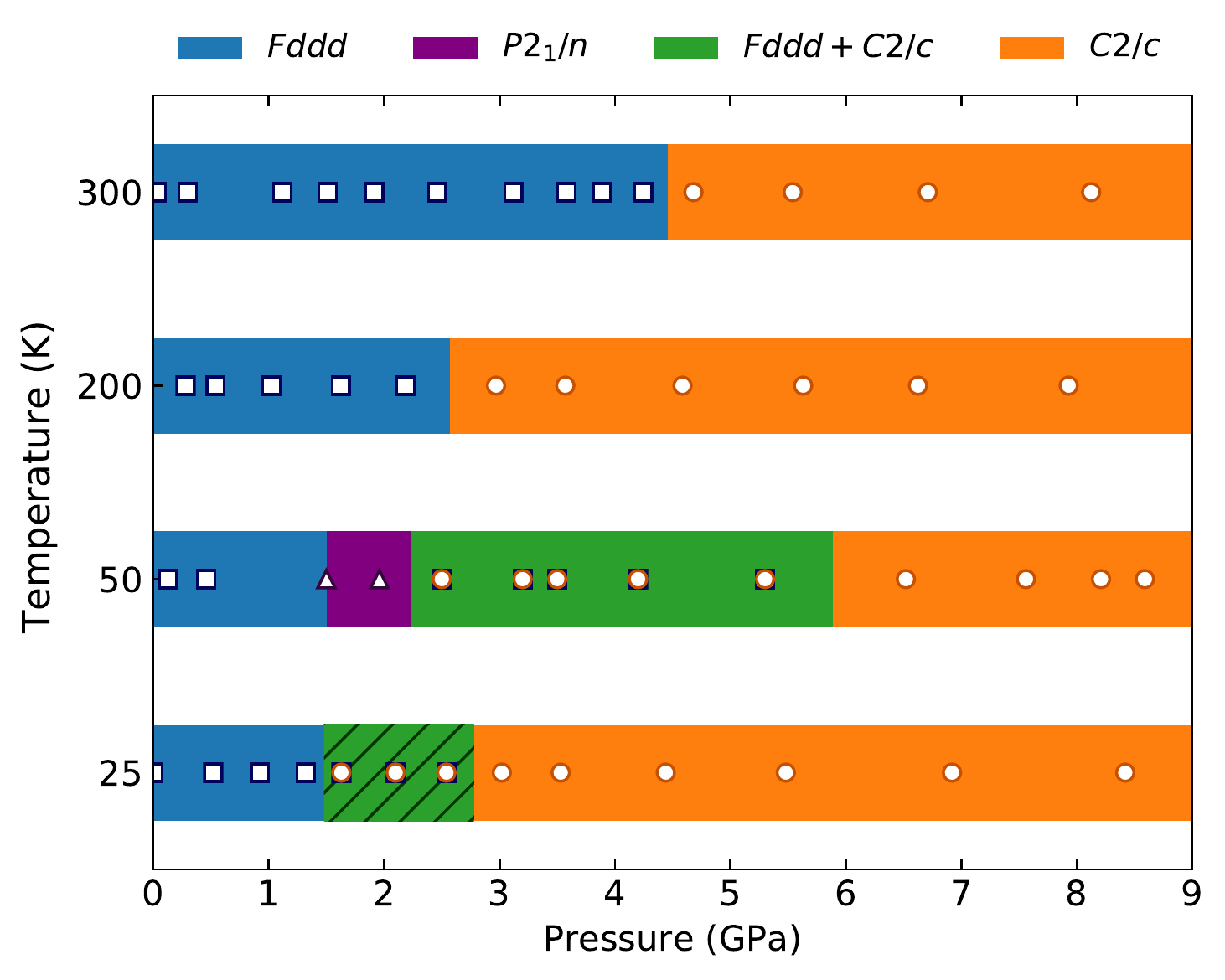}
	\caption{Identified crystal structure phases of {\Li} as a function of pressure for several temperatures. The markers represent the pressures where the XRD measurements were taken. Each temperature corresponds to an independent loading of the DAC with single crystals. We note that  an unambiguous solution using $Fddd$ and $C2/c$ space groups could not be reached for the coexistence region at $T=25$ K. While the phase coexistence region at $T=50$ K consists of dimerized $Fddd$ and $C2/c$ phases, the $Fddd$ structure in the coexistence region at $T=25$ K appears to be non-dimerized, as explained in the text and Appendix~\ref{XRD}.} 
	\label{Figure1}
\end{figure}

\begin{figure}
	\centering
	\includegraphics[width=\linewidth]{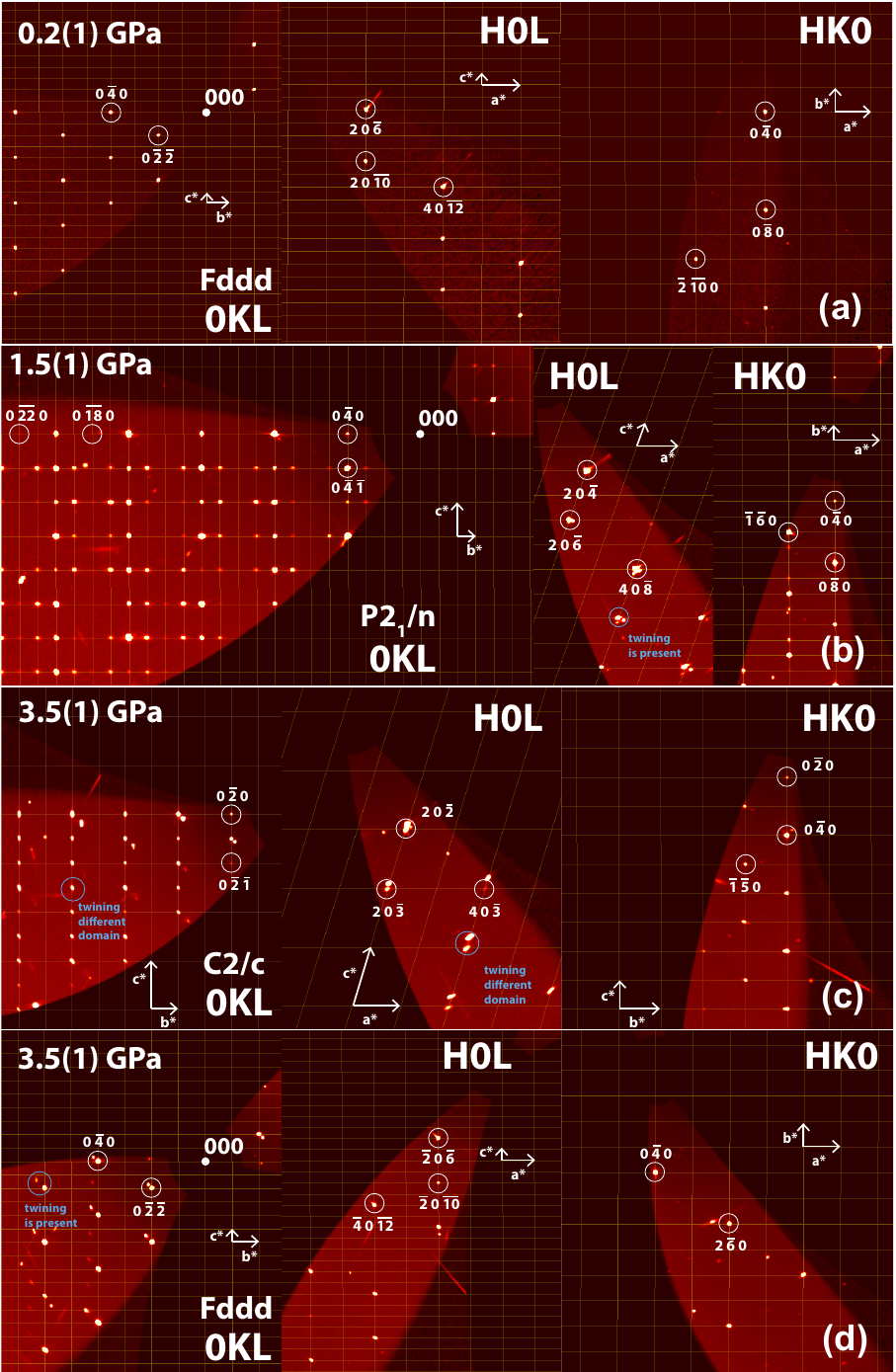}
	\caption{(Color online) Observed x-ray diffraction patterns for {\Li} at $T=50$ K shown for three different plane families for (a) the $Fddd$ orthorhombic, (b) the dimerized $P2_1/n$ monoclinic, (c) the dimerized $C2/c$ monoclinic and (d) the dimerized $Fddd$ orthorhombic structures. The $\bf{a^*}$, $\bf{b^*}$ and $\bf{c^*}$ denote the reciprocal lattice vectors of each structure. The blue circles indicate the twinning formed once the crystal undergoes the structural phase transitions. }
	\label{AP6}
\end{figure}

Compression experiments revealed a remarkably complex P-T phase diagram for {\Li}. Figure~\ref{Figure1} summarizes the identified crystal structures as a function of pressure for several temperatures. A clear suppression of the stability regime of the pure conventional $Fddd$ phase to lower pressures is observed with reduction in temperature. For $T>50$ K, the simpler $Fddd \rightarrow C2/c$ transformation is observed. Usually, such transition is accompanied by the formation of two $C2/c$ twins from a single $Fddd$ domain, with our current and previous studies~\cite{Veiga2017} indicating that this is an inherent property of the material. 

\begin{figure*}
	\centering
	\includegraphics[width=\linewidth]{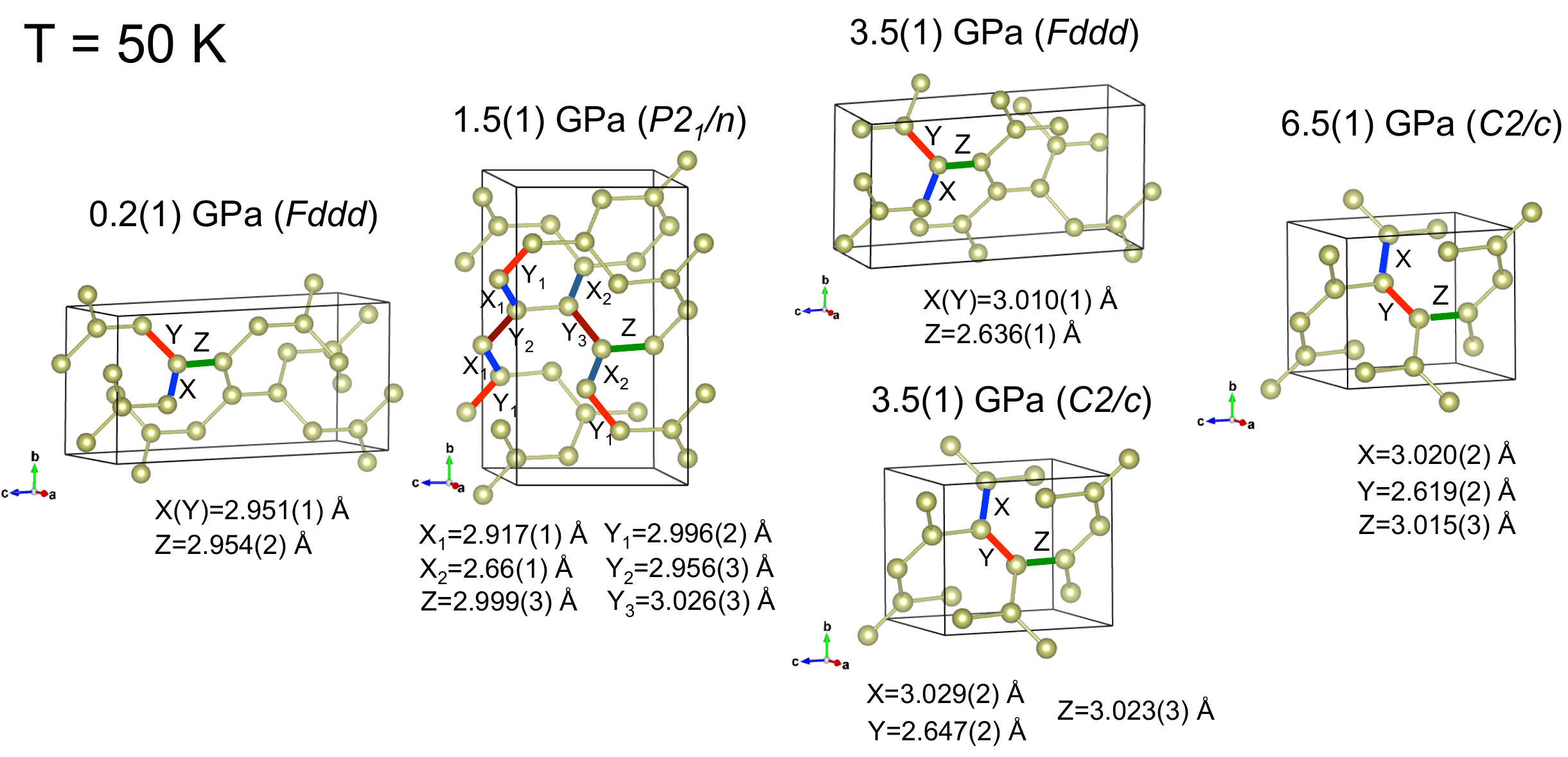}
	\caption{(Color online) Ir sublattices for all phases at $T=50$ K. The $P2_1/n$ structure is characterized by a separation of inequivalent $X$, $Y$-bonds, with $Z$-bonds nearly degenerate ($Z_1=2.998(3)$ {\AA} and $Z_2=2.999(3)$ {\AA}). A modulation in $X$ and $Y$-bonds is observed, where one of the $X$-bonds is dramatically shortened ($X_2=2.66$ {\AA}). $Fddd$ lattice in the coexistence phase is described by short Ir-Ir bonds along $Z$ direction. All the $C2/c$ structures contain shortened $Y$-bonds.}
	\label{Figure4}
\end{figure*}

For temperatures lower than 50 K, a series of structural phase transitions is found with increase of pressure, which slightly depart from the direct $Fddd \rightarrow C2/c$ transformation (Fig.~\ref{Figure1}). The XRD patterns at $T=50$ K shown in Fig.~\ref{AP6} were successfully solved using the ambient-pressure, orthorhombic $Fddd$ space group below 1.5 GPa. After that, a transition to the $P2_1/n$ phase is observed and is stable in a narrow region of pressure. Further compression transforms the $P2_1/n$ phase into a region of coexistence of $Fddd$ and $C2/c$ phases, where both structures could be independently indexed. Finally, a complete transformation to the monoclinic $C2/c$ space group is reached. Details of the identified crystal structures at $T=50$ K is shown in Fig.~\ref{Figure4} and in Ref.~\onlinecite{Note1}. 



While the $P2_1/n$ phase was not identified in the $T=25$ K isotherm, an unambiguous solution of the structure within the $Fddd$ and $C2/c$ space groups between $\sim 1.5 - 2.8$ GPa was not reached (Fig.~\ref{Figure1}). We note that the $Fddd$ structure in this coexistence region appears to be non-dimerized, opposite to what is observed for the $Fddd$ structure in the coexistence region at 50 K (Fig. \ref{Ir-Ir_25K}). Further details of the phases present in the $T=25$ K isotherm can be found in Appendix~\ref{XRD}. The new and coexisting phases at $T\leq50$ K appear in the exact pressure regime where suppression of magnetic order and strong deviation from $J_{\rm{eff}}=1/2$ state were observed~\cite{Veiga2017}. Density functional theory (DFT) calculations of the energetics of the candidate crystal structures are qualitatively consistent with our experimental results (Appendix~\ref{DFT}).

\begin{figure}
	\centering
	\includegraphics[width=\linewidth]{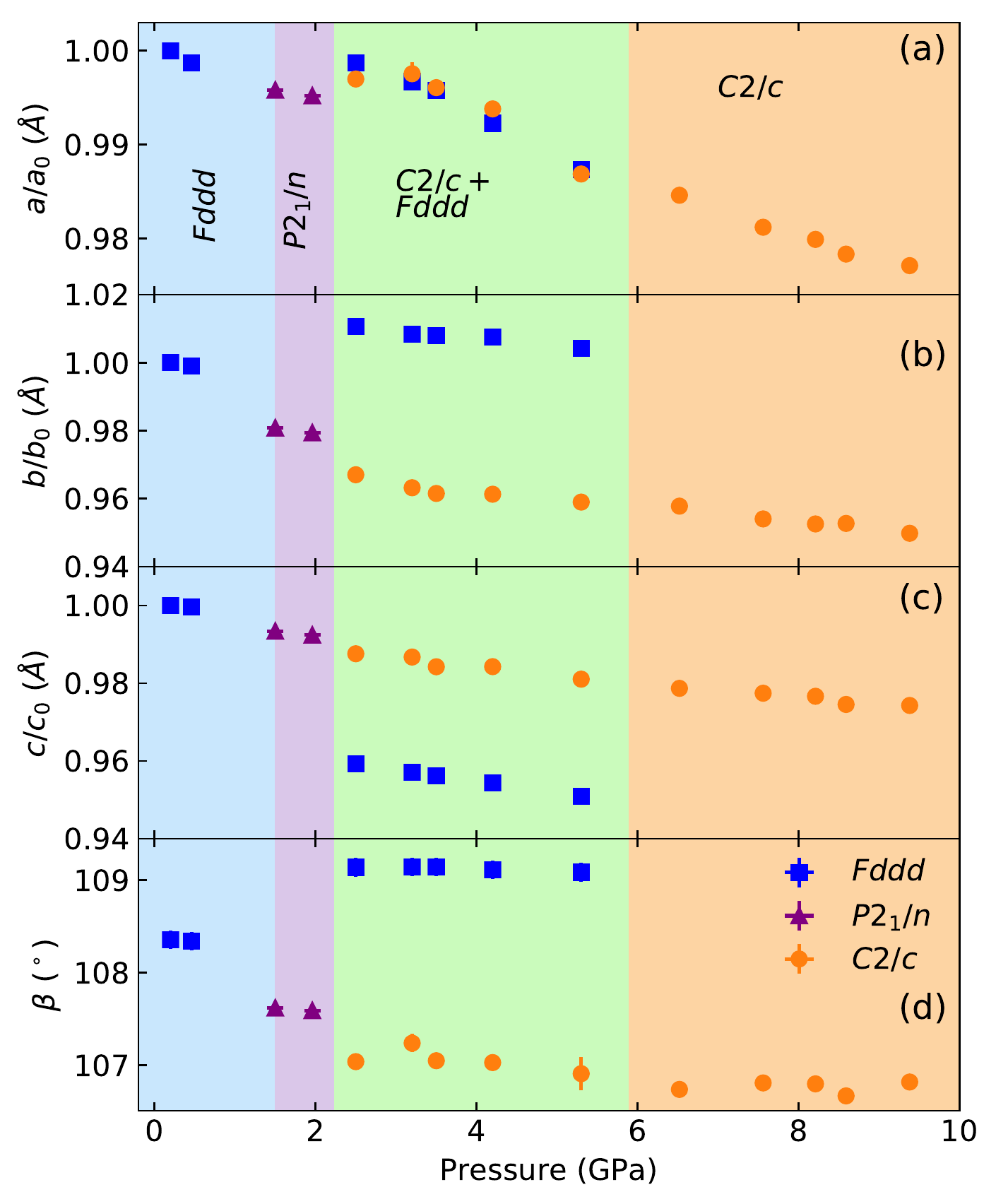}
	\caption{(Color online)(a) - (d) Lattice parameters of {\Li} at $T=50$ K as a function of pressure. At this temperature, {\Li} undergoes a series of structural phase transitions with increasing pressure, which depart from the sequence of transformation observed at 300 K. All lattice parameters were transformed to the lowest symmetry, monoclinic space group $P2_1/n$, following a group-subgroup transformation and are normalized to unity ($a_0=5.866(1)$ {\AA}, $b_0=16.7672(8)$ {\AA} and $c_0=9.3145(9)$ {\AA} at 0.2(1) GPa). If not shown, the error bar is smaller than the symbol size.}
	\label{Figure2}
\end{figure}

With the general trend of the stability of the $Fddd$ phase moving to lower pressures upon cooling, we will explore in more detail the collected isotherm at $T=50$ K. Figure~\ref{Figure2} shows the pressure dependence of the lattice parameters obtained from the analysis of the single-crystal XRD data. Additional details of the single-crystal and powder XRD analysis and results for other temperatures are presented in the Appendix~\ref{XRD}. For a better comparison between the different phases, all lattice parameters were transformed into the lowest symmetry, monoclinic space group ($P2_1/n$), following a direct group-subgroup transformation ($Fddd \rightarrow C2/c \rightarrow P2_1/n$). An anisotropic contraction of the lattice constants is seen for most phases (see Table~\ref{tab2}). Interestingly, the $a$ lattice parameter has a stronger pressure dependence than its $b$ and $c$ counterparts in the low pressure, $Fddd$ ($P<1.5$ GPa) phase. This situation changes upon entering the $P2_1/n$ phase, where the compression rate of the $b$ lattice parameter seems largest. Upon transition to the mixed phase ($P \sim 2.2$ GPa), all lattice parameters exhibit anomalies: while the $a$-axis suffers a small jump ($\sim 0.2 - 0.35 \%$) followed by a monotonic decrease with pressure, the $b$ and $c$ axes behave differently for both structures. A remarkable expansion and compression of the $b$ ($\sim3.1\%$) and $c$ ($\sim-3.3\%$) lattice parameters is observed for the $Fddd$ structure in the coexistence phase, indicating that such structure holds distinct features compared to those of the low-pressure, pure orthorhombic space group. No abrupt changes in the lattice parameters of $C2/c$ phase were observed throughout the phase boundaries and a regular contraction of the unit cell takes place up to $P=9.4$ GPa.



\begin{table}
\caption{Compression rates of the lattice parameters and volume for each identified structure at $T=50$ K. The compression rate is defined as $\frac{\frac{x-x_0}{x_0}}{p-p_0}$, where $x$($x_{0}$) is the final(initial) parameter and $p$($p_0$) is the final(initial) pressure for each phase boundary. All values are in units \%/GPa.}
\resizebox{\columnwidth}{!}{%
\begin{tabular}{ccccc}
\hline
\hline 
\
     & $\frac{\Delta a/a_0}{\Delta P}$ & $\frac{\Delta b/b_0}{\Delta P}$ & $\frac{\Delta c/c_0}{\Delta P}$ &$\frac{\Delta V/V_0}{\Delta P}$  \\
   \hline 
    
        $Fddd$ (pure)  &  -0.39(2) & -0.28(2)  & -0.12(1)& -0.83(3)  \\
        $P2_1/n$ & -0.13(3) & -0.31(1) & -0.21(2) & -0.61(4)   \\
        $Fddd$ (mixed phase)& -0.41(4)& -0.23(1) & -0.31(1) & -0.91(1)   \\
       $C2/c$ & -0.29(5) & -0.26(1) & -0.20(6) & -0.71(3)      \\
       
\hline  \hline \noalign{\smallskip}

\label{tab2}
\end{tabular}
}
\end{table}

\begin{table}[b]
\squeezetable
\caption{Comparison between the shortest ($d_{Ir}^{short}$) and average ($d_{Ir}^{av}$) Ir-Ir bond distances in {\Li} and other structures presenting dimerization. $\Delta d_{Ir}$/$d_{Ir}^{av}$ gives the measure of the variation between the shortest and longest Ir-Ir bond distances compared to the average ones. }
\resizebox{\columnwidth}{!}{%
\begin{tabular}{ccccccc}
\hline
\hline 
    Compound & Method & Temperature & Space Group & $d_{Ir}^{short}$ & $d_{Ir}^{av}$ & $\Delta d_{Ir}/d_{Ir}^{av}$ \\
   & & (K) &(\AA) & (\AA) & (\AA) & \\
    \hline
        {\Li}  &  XRD ~\cite{Takayama15} & 300 & $Fddd$ & 2.9729 & 2.9757 & 0.2\%  \\
        {\Li}  &this work & 50 & $P2_1/n$ & 2.6596 & 2.9357 & 12\%  \\
        {\Li}   & XRD~\cite{Veiga2017}& 300 & $C2/c$ & 2.662  & 2.896  & 12\%  \\
       $\alpha$-Li$_2$IrO$_3$ & XRD~\cite{Hermann2018}& 300 & $P\overline{1}$ & 2.69 &  2.903 & 11\%   \\
        $\alpha$-RuCl$_3$  &  XRD~\cite{Bastien2018}& 300 &$P\overline{1}$ & 2.86 &  3.308 & 20\%   \\
        Na$_2$IrO$_3$ &  DFT~\cite{Hu2018}& - &$P\overline{1}$ & 2.641 &  2.922 & 15\%   \\
        Li$_2$RuO$_3$ & XRD~\cite{Miura2007}  &300 & $P2_1/m$ & 2.568 & 2.887 & 18\%   \\

\hline  \hline \noalign{\smallskip}

\label{tab1}
\end{tabular}
}
\end{table}

\begin{figure}
	\centering
	\includegraphics[width=\linewidth]{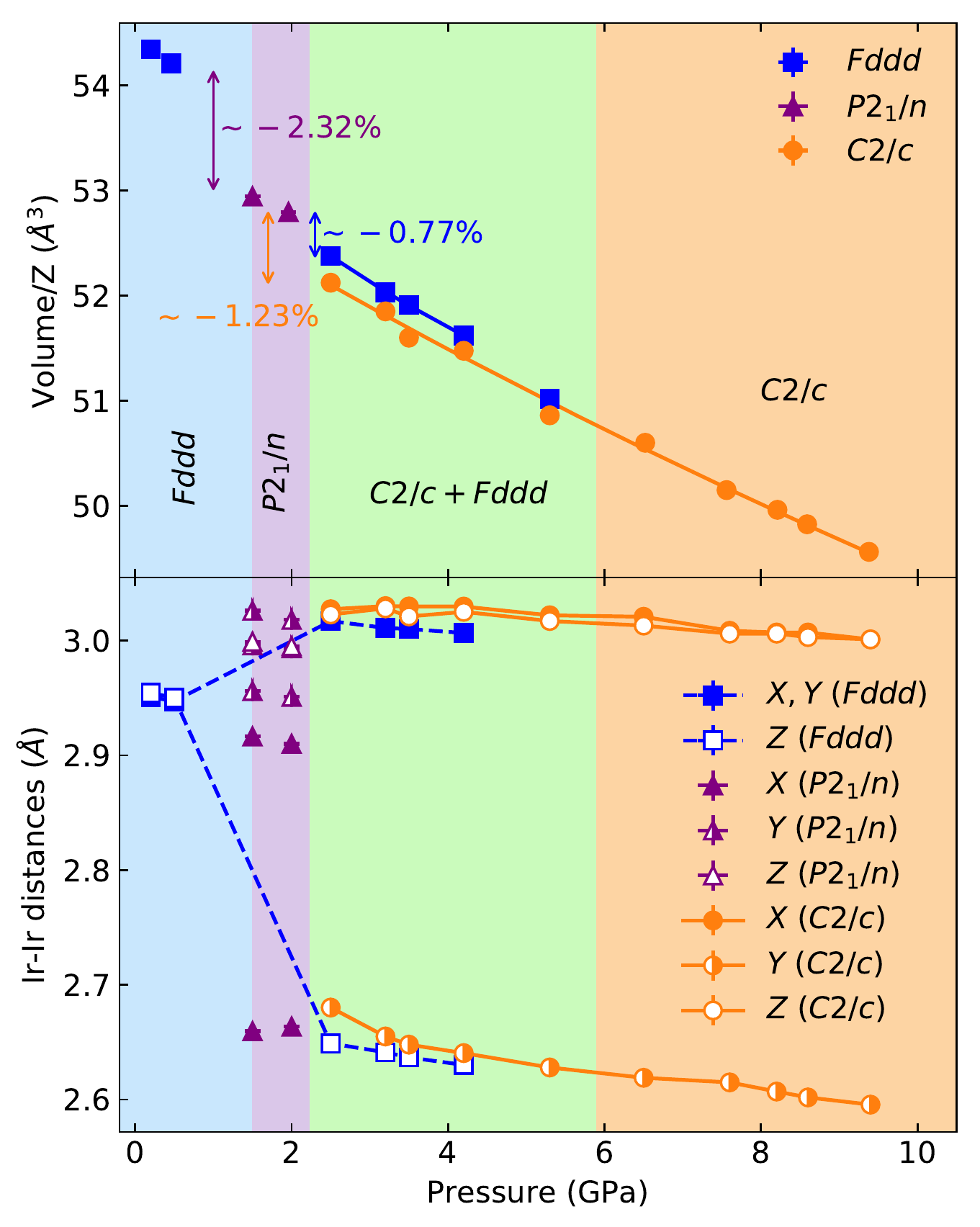}
	\caption{(Color online) (a) Pressure-volume relationship at 50 K and their fit to a second-order Birch-Murnaghan equation of state (EoS). The EoS was only determined for the high-pressure structures after the $P2_1/n$ phase ($P>2$ GPa) due to the limited data points for the low-pressure $Fddd$ and $P2_1/n$ structures. (b) Pressure dependence of the Ir-Ir distances at 50 K. The $X$-, $Y$- and $Z$-bond nomenclature of the $Fddd$ structure was kept for the $P2_1/n$ and $C2/c$ phases. If not shown, the error bar is smaller than the symbol size.}
	\label{Figure3}
\end{figure}

The evolution of volume against pressure is shown in Fig.~\ref{Figure3}(a). A small volume discontinuity (volume collapse) is observed across the phase boundaries, reaching $\sim-2.32(3)$\% ($Fddd \rightarrow P2_1/n$),  $\sim-0.77(2)$\% ($P2_1/n \rightarrow Fddd$) and $\sim-1.23(1)$\% ($P2_1/n \rightarrow C2/c$) [volume collapse size is defined as (V$_{a}$-V$_{b}$)/V$_0$, where V$_{a}$(V$_{b}$) is the volume right after(before) the transition and V$_0$ is the volume at 0.2(1) GPa]. Similar volume collapses have been observed in {\Li} at 300 K~\cite{Veiga2017} ($\sim0.7$\% at the $Fddd \rightarrow C2/c$ transition) and $\alpha$-Li$_2$IrO$_3$ ($\sim3.3$\% at the $C2/m \rightarrow P\bar{1}$ transition~\cite{Hermann2018}), and they appear to be signatures of the transition into a dimerized phase. The equation of state (EoS) for the high-pressure phases are also shown in Fig.~\ref{Figure3}(a). Fits to the Birch-Murnaghan EoS of second order~\cite{Birch1947} resulted in bulk modulus of 104(3) GPa for the $Fddd$ structure in the coexistence phase and 115(3) GPa for the $C2/c$ space group. The former value agrees well with the bulk modulus found for the $Fddd$ phase at $300$ K~\cite{Veiga2017}. The low bulk modulus of {\Li} compared to the other iridates, such as Sr$_2$IrO$_4$ (174 GPa) and $\alpha$-Li$_2$IrO$_3$ (125 GPa), reveals the relative high compressibility of this structure. Due to a limited number of data sets for the low-pressure $Fddd$ and $P2_1/n$ phases, convergence of the EoS fits could not be reached. 

A complete description of the evolution of the Ir-Ir bond distances across the phase boundaries is shown in Fig.~\ref{Figure3}(b). A key feature of the high-pressure phases is the presence of a very short Ir-Ir bond length relative to the average Ir-Ir bond distances in the structure (see Table~\ref{tab1}). The monoclinic $P2_1/n$ structure is characterized by inequivalent $X$, $Y$ and $Z$ bonds (Figure~\ref{Figure4}). At $P=1.5$ GPa, while most of $Y$ and $Z$ bonds are increased, one of the $X$ bonds is {\it dramatically shortened} to $\approx 2.66$ {\AA} as a consequence of faster contraction of the $a$ lattice parameter. This value is smaller than the Ir-Ir distance in metallic Ir ($2.714$ {\AA}). Note that in this case, one of the Ir zigzag chains shows dimerization while the other retains the Ir-Ir distances of non-dimerized structures. Interestingly, the $Fddd$ structure in the coexistence phase shows Ir-Ir dimerization of the $Z$-bonds as a result of the drastic compression of the $c$-axis parameter (Fig.~\ref{Figure3}(b)). The presence of a very short $Y$-bond is observed in the $C2/c$ phases at high pressures. Similar Ir-Ir bond changes during structural phase transitions were observed experimentally and theoretically for other honeycomb compounds at high-pressures ($\alpha$-Li$_2$IrO$_3$ at 3.8 GPa~\cite{Hermann2018}, $\alpha$-RuCl$_3$ at $\approx 0.2$ GPa~\cite{Bastien2018} and Na$_2$IrO$_3$ at $\approx 36$ GPa~\cite{Hu2018}, see Table~\ref{tab1} for other dimerized compounds), indicating that honeycomb-based ruthenates and iridates featuring edge-sharing octahedra are unstable towards dimerization.

\subsection{X-ray absorption near edge structure}

\begin{figure}
	\centering
	\includegraphics[width=\linewidth]{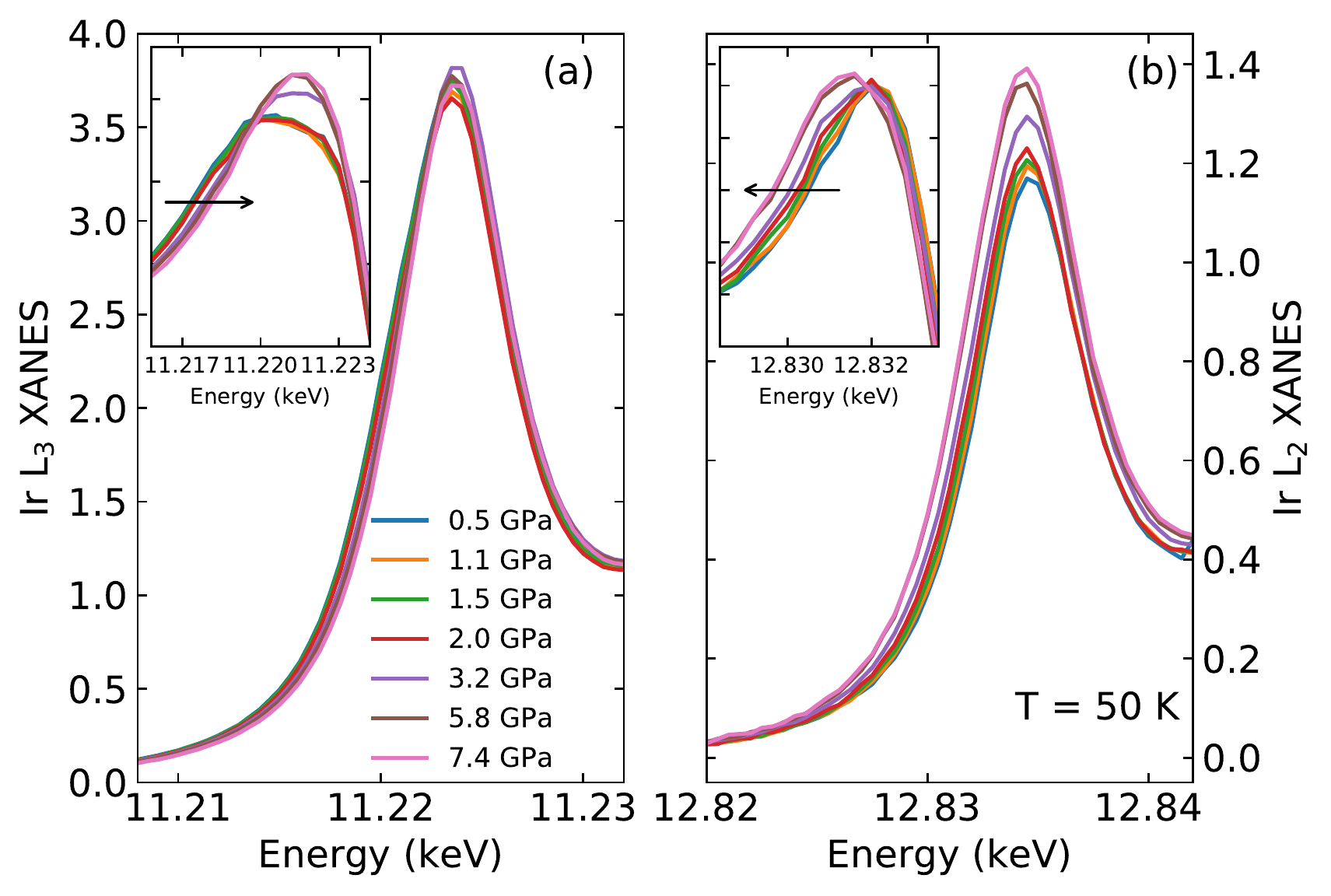}
	\caption{(Color online) (a,b) Ir $L_{2,3}$ XANES data at $T=50$ K for selected pressures. The insets show the first derivative of the XANES spectra and reveal the opposite energy shift behavior of the leading absorption edge as a function of pressure for each absorption edge.}
	\label{Figure5}
\end{figure}

\begin{figure}
	\centering
	\includegraphics[width=\linewidth]{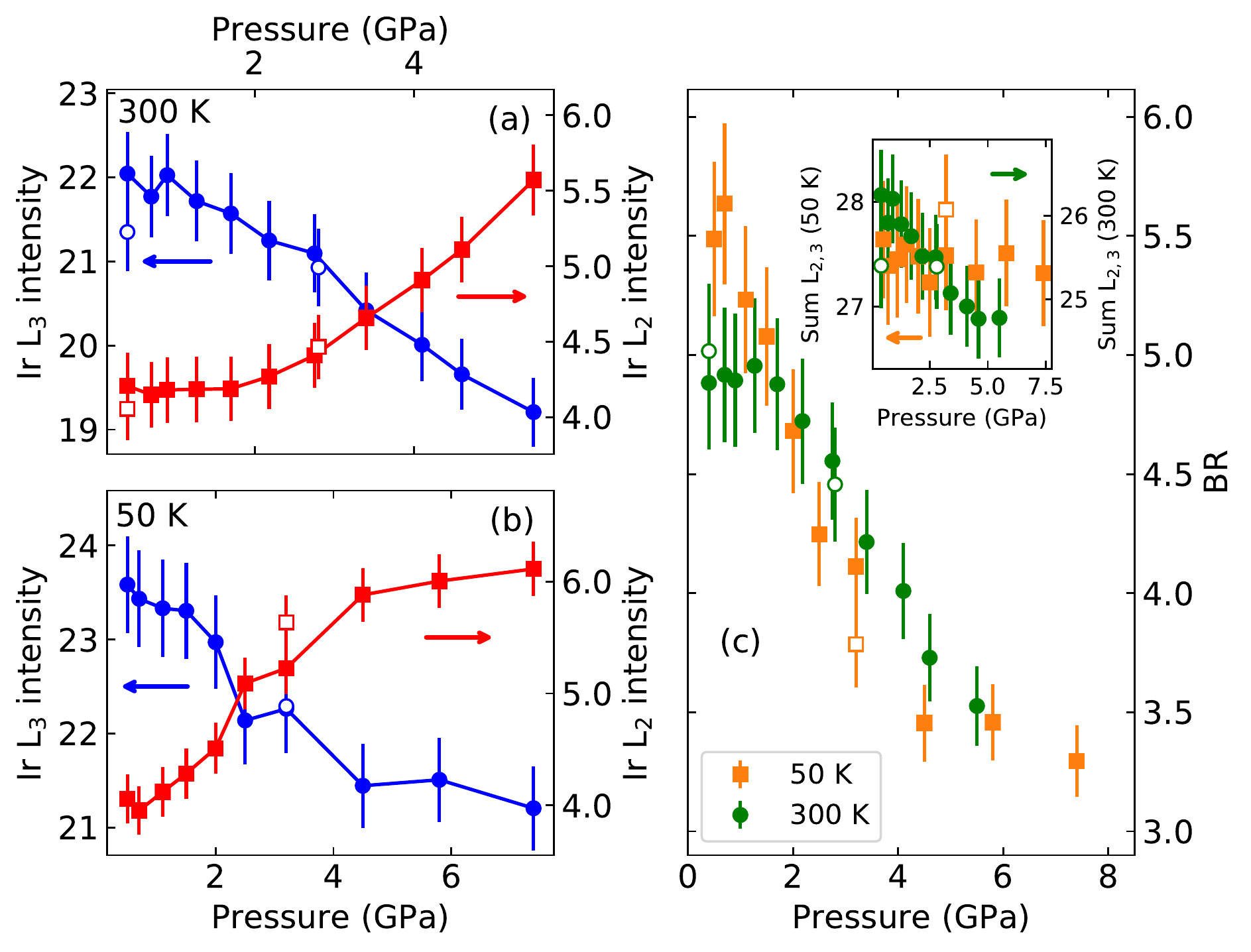}
	\caption{(Color online) Ir $L_{2,3}$ XANES integrated intensities as a function of pressure at (a) $300$ K and (b) $50$ K. (c) Pressure dependence of the branching ratio at $50$ K and $300$ K. The inset shows the sum of the Ir $L_{2,3}$ integrated intensities for each temperature, which is proportional to the number of Ir $5d$ holes. The open markers represent the data upon pressure release. }
	\label{Figure6}
\end{figure}

X-ray absorption measurements at the Ir $L_{2,3}$ edges as a function of pressure were performed at $T=50$ K and 300 K. Figures~\ref{Figure5}(a), (b), main panels, display the XANES spectra for selected pressures at $T=50$ K. The branching ratio (BR), which is defined as the white-line intensity ratio at $L_{2, 3}$ edges, $I_{L_3}/I_{L_2}$, is proportional to the expectation value of the angular part of the spin-orbit interaction $\langle {\bf L} \cdot {\bf S} \rangle$, through BR = $(2+r) / (1-r)$, with $r= \langle {\bf L} \cdot {\bf S} \rangle / n_h$ and $n_h$ the number of holes in the $5d$ states \cite{Laan88}. Figures~\ref{Figure6}(a), (b) show the integrated intensities at the Ir $L_{2,3}$ edges as a function of pressure for each temperature and Fig.~\ref{Figure6}(c), main panel, displays the pressure dependence of the BR for both temperatures. 

At very low pressures, the measured BR (BR$_{50K}=5.5(3)$, BR$_{300K}=4.9(3)$) strongly deviates from the statistical value of 2 and indicates the presence of strong spin-orbit interaction and proximity to a $J_{\rm{eff}}=1/2$ ground state~\cite{Laguna-Marco10, Haskel12, Laguna-Marco15, Clancy12}. Upon further compression, the BR at $T=300$ K is essentially constant to at least 2 GPa, after which it monotonically decreases and reduces by $\sim30\%$ of the initial value at 5.5 GPa, consistent with the opposite behavior displayed by the Ir $L_{3}$ and $L_{2}$ integrated intensities in Fig.~\ref{Figure6}(a). The midpoint drop around $P\sim3.5$ GPa is in proximity to the structural phase transition observed at this temperature ($\sim4.4(2)$ GPa). On the other hand, the pressure-dependent BR at $T=50$ K decreases steadily between $P\sim0.7-4.5$ GPa, above which assumes a constant value of $\sim3.4$. The midpoint drop at $P\sim2.6$ GPa is also close to the onset of structural phase transition observed at $T=50$ K ($P_S\sim1.5$ GPa) and agrees with the trend showed in Fig.~\ref{Figure6}(b).

We note that our previous study of the BR response to pressure also shows a BR drop that is shifted to higher pressures at 300 K relative to 5 K~\cite{Veiga2017}. However, as also found here, the BR drop at 300 K clearly starts ahead of the structural transition. The new measurements taken with NPDs and with a higher density of data points allow us to confirm a clear shift between 300 K and 50 K data sets in the threshold pressure above which the BR decreases. The shift appears to correlate with the shift in the critical pressure for the structural transformation although the correspondence is not obvious. A more clear correspondence is found in the response of the leading absorption edge, as discussed below.


That the onset of a gradual reduction in BR takes place at pressures lower than those required to drive the structural phase transition both at 50 K and 300 K may be indicative of early onset of local distortions that are not captured in the XRD measurement, although changes in inter-site hybridization with pressure can also be at play. We note that the width of the BR transition is significantly larger than expected from pressure gradients in the mineral oil pressure medium ($<0.5$ GPa), particularly at 300 K~\cite{Klotz2009}. In addition, other explanations for the BR drop, such as charge transfer from oxygen to Ir sites, can be ruled out by our data as revealed by the opposite energy shifts of $L_3$ and $L_2$ absorption edges with pressure (see insets of Fig.~\ref{Figure5} for the 50 K and Fig.~\ref{AP4} in Appendix~\ref{XANES} for the 300 K data). This is further confirmed by the rather constant or small changes (reduction of $5\%$) in the number of holes observed at $50$ K and 300 K, respectively (inset of Fig.~\ref{Figure6}(c)). The energy shift of the leading absorption edges with pressure (Fig.~\ref{AP5}) shows a stronger correlation with the structural transition compared to the more gradual change in BR leading support to the idea that the BR may also be affected by other mechanisms, e.g., increased hybridization upon compression. Nonetheless, our results indicate that the suppression of BR and collapse of the $J_{\rm{eff}}=1/2$ state is directly associated with an instability towards Ir$_2$ dimer formation in this system under high pressures.



\section{Discussion}
Our XRD data clearly indicate that the structural phase transition from an orthorhombic to a dimerized monoclinic structure is suppressed to lower pressures at low temperatures ($P_S\sim 4.4$ GPa at 300 K to $P_S\sim1.5$ GPa at 50 K) and therefore constitutes the main cause for the collapse of magnetic order and departure from the $J_{\rm{eff}}=1/2$ ground state found in our previous studies~\cite{Veiga2017}. In particular, our XRD findings reveal that all the identified phases follow the crystallographic group-subgroup transformation (such as martensitic). Such transformations do not require as much energy as, for example, reconstructive transformations (thermally activated diffusion processes) and, therefore, the individual phases can be very close in energy. Thus, any additional contribution to the total energy, such as from strain, would shift the phase equilibrium towards these energetically close structures. In addition, our data indicates that in the regime of strongly suppressed kinetics, the material undergoes an inherent twinning (acting as an additional strain), which is intrinsic to the system and is also present in powder XRD patterns collected at high pressures and low temperatures (see Appendix~\ref{XRD}).

A key observation of our XRD results at $T=50$ K is the different ways in which the lattice parameters contract across the identified phases. In the low pressure, orthorhombic $Fddd$ phase, the $a$ lattice parameter contracts at a faster rate than its $b$ and $c$ counterparts. This leads to a transformation to the $P2_1/n$ structure, where one of the inequivalent $X$-bonds is \emph{drastically} shortened, indicating formation of Ir$_2$ dimers along this direction in half of the zigzag chains forming the structure. Stronger contraction of the $b$ lattice parameter, on the other hand, is found when the $P2_1/n$ phase is compressed, leading to a conversion to the $C2/c$ structure with dimerized Ir-Ir bonds along $Y$ direction. Note that a rather significant contraction of the $c$ lattice parameter is also observed in the $P2_1/n$ phase, a fact that could explain the unexpected transition to the $Fddd$ structure in a mixed phase, with presence of shortened $Z$-bonds. While the latter was not detected in any of the experiments conducted in the past~\cite{Veiga2017, Takayama2019}, the presence of dimerized $Z$-bonds was predicted theoretically for structure optimizations in absence of SO coupling or Coulomb interaction~\cite{Kim16}. Although our results point towards a dimerization of all Ir-Ir bond types, a preferable dimerization of the Ir-Ir $X$, $Y$-bonds forming the zig-zag chains is observed as a consequence of the significant contraction of the $a$ and $b$ lattice parameters for almost all the structures. We note that our DFT calculations of the energetics of the different crystal structures captures the $Fddd \rightarrow C2/c$ transition, with the $C2/c$ phase containing dimerized $Y$-bonds (see Appendix~\ref{DFT} for further information). The calculations also show that competing dimerized phases with different short bond selection may be at play in compressed {\Li}.

Our x-ray absorption spectroscopy data reveals that the reconstruction of the electronic ground state is correlated with the onset of dimerized phases. While pressure gradients in the mineral oil pressure medium will broaden the pressure range of the BR transition, these gradients ($\sim0.5$ GPa) are expected to be much smaller than the observed width of the BR transition. Since the BR is a local probe of electronic states and it does not rely on long-range order, it is possible that local dimer formation (either static or dynamic) or other local structural distortions appear before dimers "condense" at the various structural phase transition. Also, enhanced inter-site hybridization cannot be ruled out as possible contributor for the gradual reduction in BR. It should be noted that in the intermediate $P2_1/n$ phase at $50$ K, a potential coexistence of dimerized nonmagnetic and localized $J_{\rm{eff}} = 1/2$-like spins in the Ir zigzag chains (Fig.~\ref{Figure4}) could also be contributing to the slow reduction in BR, although the extent of such is overshadowed by the other contributions. Nevertheless, our results suggest that the formation of the dimerized phase leads to a collapse of the $J_{\rm{eff}}=1/2$ state and possible formation of non-magnetic singlet state, all the while the system remaining insulating. Indeed, our temperature- and field-dependent XMCD data~\cite{Veiga2017} shows that only a very small paramagnetic response is present even at low temperatures (5 K) and applied field (4 T), which is consistent with formation of a spin-singlet dimer state. Moreover, recent resonant inelastic x-ray scattering (RIXS) measurements together with electronic structure calculations~\cite{Takayama2019} show a drastic reconstruction of the electronic ground state associated with this dimerization, which stabilizes a bonding molecular-orbital state with predominant $d_{zx}$-orbital character, which is inconsistent with the equal superposition of $d_{xy}$, $d_{yz}$, $d_{zx}$ orbitals present in the $J_{\rm{eff}}=1/2$ wave function.


We now turn to the implications of the dimerized phase on the magnetic state of {\Li} at high pressures. The collapse of magnetic order seen by XMCD~\cite{Takayama15, Veiga2017} and more recently by magnetometry and $\mu$SR experiments~\cite{Majumder2018} in the $\sim1.5-2$ GPa range at low temperature was interpreted as a purely electronic effect based on structural measurements carried out at room temperature. Our single crystal and powder XRD experiments, on the other hand, confirm that a structural phase transition takes place at $P_S \sim 1.5$ GPa at \emph{low temperatures} and, therefore, formation of Ir$_2$ dimers is the leading mechanism driving a magnetically disordered ground state at high pressures. Local dimer fluctuations and/or coexistence of non-dimerized and dimerized phases may be responsible for the coexistence of dynamic and frozen spin states seen in  $\mu$SR data~\cite{Majumder2018}.

The instability towards formation of Ir$_2$ dimers reveals that a subtle interplay between magnetism, electron correlation, spin-orbit coupling, and covalent bonding is at play in the hyperhoneycomb iridate {\Li}. At ambient pressure, the presence of a single hole in the Ir $t_{2g}$ manifold combined with a strong spin-orbit coupling, reduces the propensity to dimerization and long-range magnetic order emerges. Pressure increases the overlap of the orbitals by bringing Ir ions closer together, thus weakening the tendency to magnetism and favoring the formation of Ir$_2$ molecular orbitals. It should be noted that dimerization was also identified in the 2D-honeycomb iridate $\alpha$-Li$_2$IrO$_3$ under similar pressures~\cite{Hermann2018}, with calculations indicating a concomitant magnetic collapse at the same pressure. Moreover, other two-dimensional $4d$ honeycomb structures, such as Li$_2$RuO$_3$~\cite{Miura2007} and $\alpha$-MoCl$_3$~\cite{McGuire17}, are dimerized at ambient pressure while in $\alpha$-RuCl$_3$, a very small pressure of $\sim0.2$ GPa is required to form Ru$_2$ dimers~\cite{Bastien2018}, suggesting that dimerization is rather a common feature in these $4d$/$5d$ layered honeycomb systems.   

Elucidating if instability to dimer formation is also relevant for the collapse of magnetic order in other 3D-honeycomb structures is important to estabilish a universal picture of competing interactions in the honeycomb-based materials. In fact, high pressure studies of $\gamma$-Li$_2$IrO$_3$~\cite{Breznay17} reveal a pressure collapse of the magnetic ordering at $\sim1.5$ GPa, the same pressure where the XMCD, magnetometry and $\mu$SR experiments on {\Li} show breakdown of the magnetic state~\cite{Takayama15, Veiga2017,Majumder2018}. However, no clear signatures of structural dimerization were found on the former polytype, where a continuous reduction of the unit-cell volume was observed to at least $\sim3.3$ GPa. We note that such result was inferred from determination of the $a$ lattice parameter through the measurement of a single structural Bragg peak. A more detailed structural investigation of $\gamma$-Li$_2$IrO$_3$ across the pressure-temperature phase diagram is lacking to date. Nonetheless, considering the similarities of the magnetic and structural ground states between these two polytypes, it is expected that $\gamma$-Li$_2$IrO$_3$ would also dimerize at the same pressure where collapse of magnetism is observed. 

\section{Concluding remarks}

In this work we studied the pressure-temperature phase diagram of {\Li} with a focus on the low temperature region. Our XRD results clearly indicate that the critical pressure of the structural phase transition is suppressed to $P_S\sim1.5$ GPa at low temperatures ($T\leq50$ K), the same pressure where the collapse of magnetic order in {\Li} is observed. A series of structural phase transitions including new and coexistence phases was found with increase of pressure, which departs from the direct $Fddd \rightarrow C2/c$ transformation observed at $T=300$ K. The high-pressure structures are characterized by the formation of Ir$_2$ dimers in most of the Ir-Ir bond directions. However, a preferable dimerization of the Ir-Ir $X$, $Y$-bonds forming the zig-zag chains is observed as a consequence of the significant contraction of $a$ and $b$ lattice parameters for most of the structures. Reinvestigation of the electronic structure by means of x-ray absorption spectroscopy indicates that dimerization is directly related to the breakdown of the $J_{\rm{eff}}=1/2$ state at high pressures. Dimerization under high pressures also points to a potentially universal competition between spin-orbit coupling, magnetism, and formation of molecular orbitals in Ir$_2$ dimers in the honeycomb-based iridates and confirms the subtle nature of the low-energy physics of these systems.

\begin{acknowledgements}

This work is supported by the UK Engineering and Physical Sciences Research Council (Grants No. EP/N027671/1 and No. EP/N034694/1). Work at Argonne is supported by the US Department of Energy, Office of Science, Office of Basic Energy Sciences, under Contract No. DE-AC- 02-06CH11357. H. P. is supported by the US Dept. of Energy, Office of Science, Basic Energy Sciences, Materials Sciences and Engineering Division. We thank D. Robinson for the help with the setup of the powder XRD experiment at 6-ID-D beam line and GSECARS for use of laser drilling facilities. We also thank PRIUS for providing nanopolycrystalline diamond anvils. Part of this research was carried out at PETRA III at DESY, a member of Helmholtz Association (HGF). 

\end{acknowledgements}

\appendix
\section{Powder and single crystal XRD measurements at high pressures and low temperatures}
\label{XRD}

The high pressure powder XRD experiments were performed at beam line 6-ID-D of the Advanced Photon Source using a membrane driven diamond anvil cell (DAC) with large angular aperture (2$\theta=60^{\circ}$) using Bohler-Almax diamond anvils with culet size of 800 $\mathrm{\mu m}$. A 400 $\rm{\mu m}$ hole in a stainless steel gasket pre-indented to 91 $\rm{\mu m}$ was filled with {\Li} powders together with Au and 4:1 methanol:ethanol as the pressure-transmitting medium. The measurements were carried out at $T=11$ K using a closed-cycle cryostat. The x-ray energy was tuned to 51.3 keV ($\lambda=0.242$ {\AA}) and the two-dimensional (2D) XRD patterns were recorded with a MAR345 image plate and converted into 1D plots using the FIT2D software \cite{Hammersley96}. 

\begin{figure}[t]
	\centering
	\includegraphics[width=\linewidth]{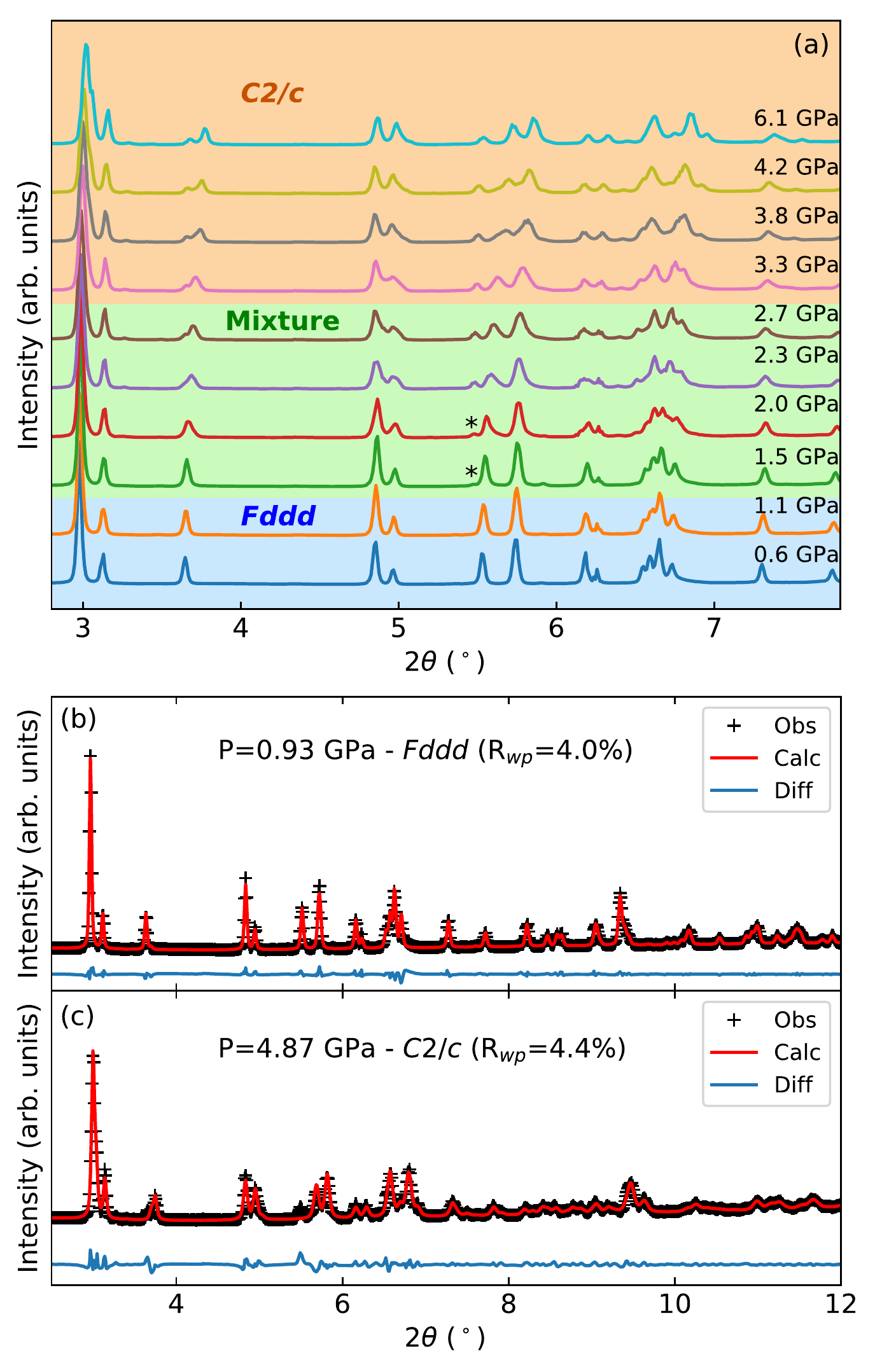}
	\caption{(Color online) (a) Angle dispersive x-ray powder diffraction of {\Li} as a function of pressure at $T=11$ K. At $\sim1.5$ GPa, a new weak peak around $2\theta=5.45^\circ$ appears (marked by *), followed by more new peaks after $\sim2$ GPa. Refinements of the diffraction patterns indicate a mixture of phases between 1.5 - 3.3 GPa. After 3.3 GPa, the XRD patterns could be refined within the pure $C2/c$ space group. (b, c) LeBail refinements using pure $Fddd$ and $C2/c$ structures.}
	\label{AP1}
\end{figure}

\begin{figure*}
	\centering
	\includegraphics[width=\linewidth]{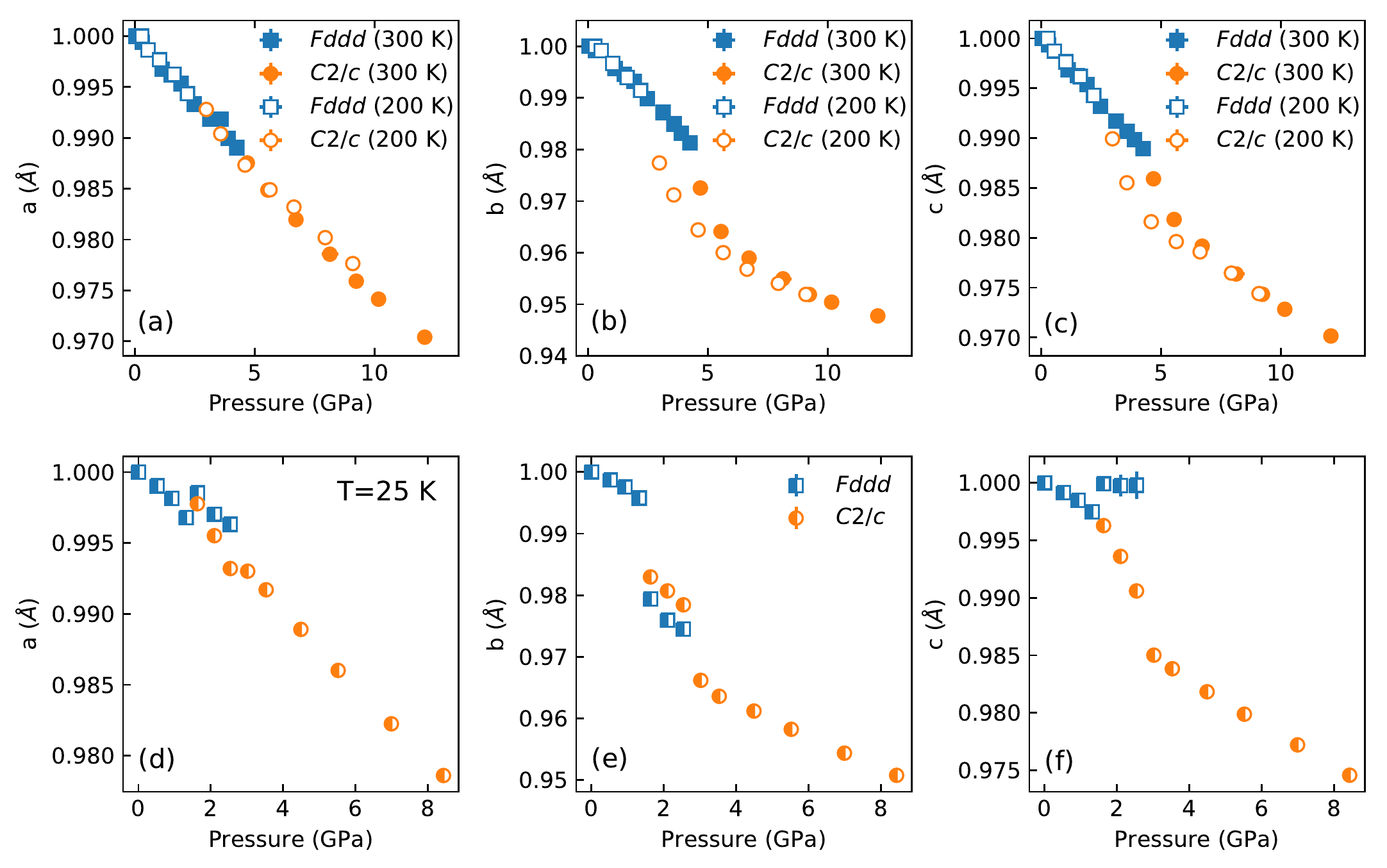}
	\caption{(Color online) (a-c) Pressure dependence of lattice parameters extracted from refinements of the single crystal XRD patterns at $T=300$ K (filled symbols) and $T=200$ K (open symbols). Note that the critical pressure of the structural phase transition is suppressed upon temperature decrease, going from $P_S\sim4.4$ GPa at $T=300$ K to $P_S\sim2.6$ GPa at $T=200$ K. (d-f)  Pressure dependence of lattice parameters at $T=25$ K. At $\sim1.5$ GPa, a structural phase transition from the orthorhombic $Fddd$ to a mixture of phases ($Fddd+C2/c$) takes place. For a better comparison between the different phases, all lattice parameters were transformed into the lowest symmetry, monoclinic space group ($C2/c$), following a direct group-subgroup transformation ($Fddd \rightarrow C2/c$).}
	\label{AP2}
\end{figure*}

\begin{table}[t!]
\caption{Compression rates of the lattice parameters and volume for each identified structure at 300 K, 200 K and 25 K. All values are in units \%/GPa.}
\resizebox{\columnwidth}{!}{%
\begin{tabular}{cccccc}
\hline
\hline 
\
  Temperature &Structure   & $\frac{\Delta a/a_0}{\Delta P}$ & $\frac{\Delta b/b_0}{\Delta P}$ & $\frac{\Delta c/c_0}{\Delta P}$ & $\frac{\Delta V/V_0}{\Delta P}$   \\
   \hline 
    
      300 K & $Fddd$  &  -0.26(1) & -0.44(1)  & -0.26(3)& -0.95(2) \\
       &$C2/c$ & -0.23(4) & -0.34(1) & -0.22(5)& -0.74(7)  \\
    \hline 
        200 K & $Fddd$  &  -0.14(1) & -0.20(6)  & -0.14(1) &-0.47(5)\\
       &$C2/c$ & -0.25(1) & -0.43(9) & -0.26(1) & -0.84(8)  \\
       \hline 
       25 K & $Fddd$ (pure)  & -0.24(1)  & -0.32(1)  & -0.19(2) &-0.75(1)\\
       &$Fddd$ (mixed) &-0.24(4)  & -0.55(6) & -0.01(1) & -0.8(1)  \\
       &$C2/c$ (mixed) &-0.50(3)  & -0.51(4) & -0.63(5) & -1.44(5)  \\
         &$C2/c$ (pure) &-0.26(1)  & -0.52(2) & -0.23(1) & -0.95(2)  \\

\hline  \hline \noalign{\smallskip}

\label{tab3}
\end{tabular}
}
\end{table}

First analysis of the powder XRD patterns were performed using LeBail method as implemented in the GSAS/EXPGUI program \cite{Larson00, Toby01}. The powder XRD patterns were successfully fitted within the orthorhombic $Fddd$ space group up to $\sim1.4$ GPa (Figure~\ref{AP1}(b)). New reflection could be identified after $\sim1.5$ GPa, signaling a structural phase transition (see Figure~\ref{AP1}(a)). Attempts to fit the new phase within $P2_1/n$ space group resulted in slightly better R-factors ($R_{wp}=4.7\%$) when compared to $C2/c$ structure ($R_{wp}=5.8\%$), a fact related to the enormous amount of reflections allowed by the low symmetry of the former space group. Since LeBail method imposes no constraints on the diffracted intensities, attempts to use Rietveld refinement were performed to determine the presence of phase coexistence after 1.5 GPa. Using the structures generated by refinements of the single crystal XRD data at low temperatures ($50$ K and 25 K), analysis of the powder XRD patterns indicate a potential coexistence of $Fddd$ and $P2_1/n$ phases up to $\sim2.3$ GPa. Between 2.3 and $\sim3.3$ GPa, while the results suggest presence of the $Fddd$ structure, the exact determination of the monoclinic phases ($C2/c$ or $P2_1/n$) was not possible. Above $\sim3.3$ GPa, either Rietveld or LeBail methods yield better refinements using $C2/c$ space group (average LeBail $R_{wp}\sim4.9\%$, see Figure~\ref{AP1}(c)). In essence, analyses of the powder XRD indicate: (1) a structural phase transition takes place at $P_S\sim 1.5$ GPa at $T=11$ K and (2) coexistence of phases, including $Fddd$, $C2/c$ and/or $P2_1/n$, is also present in the powder data for a considerable range of pressure. 


Additional single-crystal XRD patterns were collected at $T=300$ K, $200$ K and $25$ K. The evolution of the lattice parameters are shown in Figure~\ref{AP2}. For $T=300$ K and 200 K, a structural phase transition from the orthorhombic $Fddd$ to the monoclinic $C2/c$ space group takes place without the presence of mixed phases. Upon temperature decrease, the critical pressure of the transition is suppressed from $\sim4.4(2)$ GPa ($T=300$ K) to $\sim2.6(4)$ GPa ($T=200$ K). An overview of the lattice compression rates is presented in Table~\ref{tab3}. An anisotropic contraction of the lattice parameters, with $b$ parameter contracting at faster rate than its $a$ and $c$ counterparts is seen for all phases at $300$ K and 200 K. Compression rate values for the $Fddd$ phase agrees well with those reported in Ref.~\onlinecite{Veiga2017}.    

\begin{figure}
	\centering
	\includegraphics[width=\linewidth]{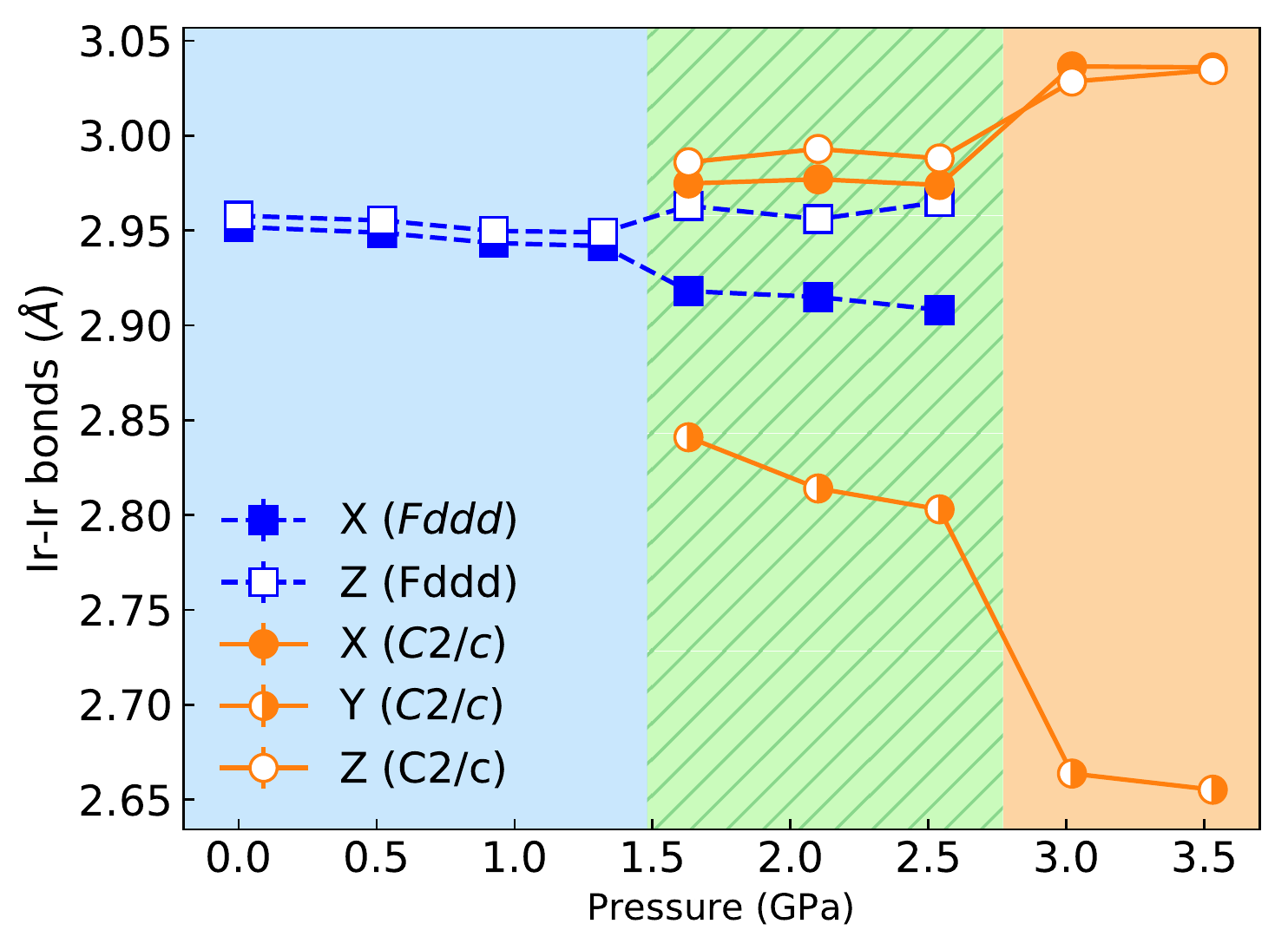}
	\caption{(Color online) Pressure dependence of the Ir-Ir distances for the identified structures at $T=25$ K. If not shown, the error bar is smaller than the symbol size.}
	\label{Ir-Ir_25K}
\end{figure}

Inspection of the lattice parameters evolution as a function of pressure at $T=25$ K reveals that the structural phase transition takes place at a reduced pressure of $P_S\sim1.5(2)$ GPa (see Fig.~\ref{AP2}(d-f)). Transition to a mixed phase (slightly distorted $Fddd$ transforming to either $Fddd+C2/c$ or double $C2/c$ domains) is observed before a complete transformation to a pure $C2/c$ structure at $\sim3$ GPa. No signatures of the monoclinic $P2_1/n$ space group were found in the entire pressure range measured. We note that due to the strong overlap of the diffraction peaks of the $Fddd$ and $C2/c$ phases, an unambiguous interpretation of the phase assemblage between $\sim 1.5 - 2.8$ GPa was not possible. Further inspection of the identified structures at $T=25$ K reveals that the Ir-Ir distances in the $Fddd$ structure in the coexistence region are longer than the ones found for the same structure in the mixed phase at $50$ K (Fig. \ref{Ir-Ir_25K}). While this structure appears to be non-dimerized, the difference between $X(Y)$- and $Z$-bonds is more pronounced than the one found in the low pressure $Fddd$ structure ($P<1.5$ GPa). We also note that the $Y$-bond in the $C2/c$ phase in the coexistence region at 25 K is longer ($\sim2.84$ {\AA}) compared to the value found for the dimerized $C2/c$ phase at other temperatures ($\sim2.66$ {\AA}), a fact that may be related to the difficulty in finding a robust structural solution in this particular region. Good R-factor values ($R\sim3.5\%$) were found for refinements using only $C2/c$ space group after $\sim 3$ GPa and are similar to the R-factors ($R\sim2.5-3.5\%$) found in the low pressure, $Fddd$ phase.                   


\section{Additional x-ray absorption spectroscopy results}
\label{XANES}

\begin{figure}
\centering
	\includegraphics[width=\linewidth]{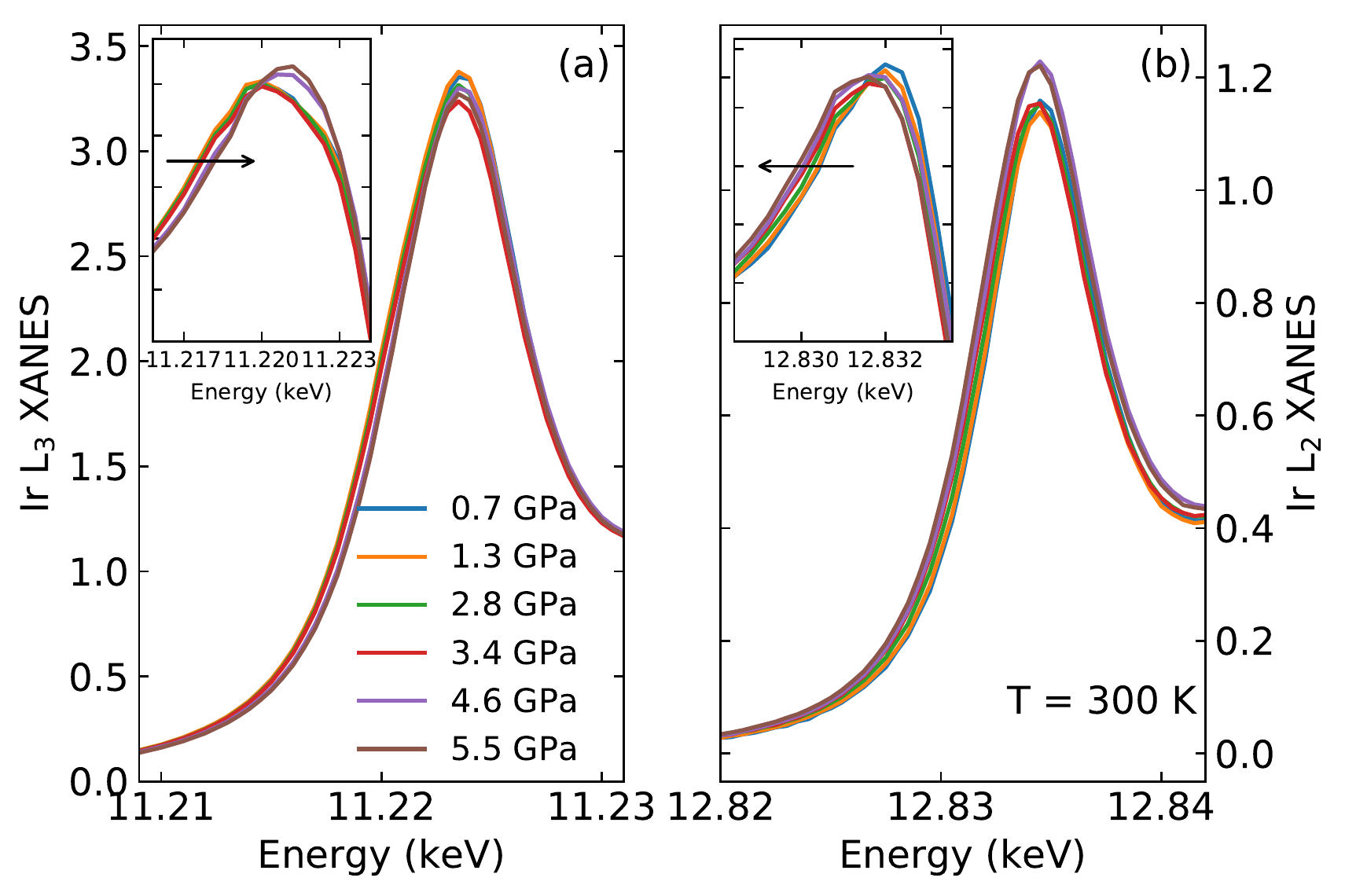}
	\caption{(Color online) (a,b) Ir $L_{2,3}$ XANES data at $T=300$ K for selected pressures. The insets show the first derivative of the XANES spectra and reveal the opposite energy shift behavior of the leading absorption edge as a function of pressure for each absorption edge.} 
	\label{AP4}
\end{figure}

\begin{figure}
\centering
	\includegraphics[width=\linewidth]{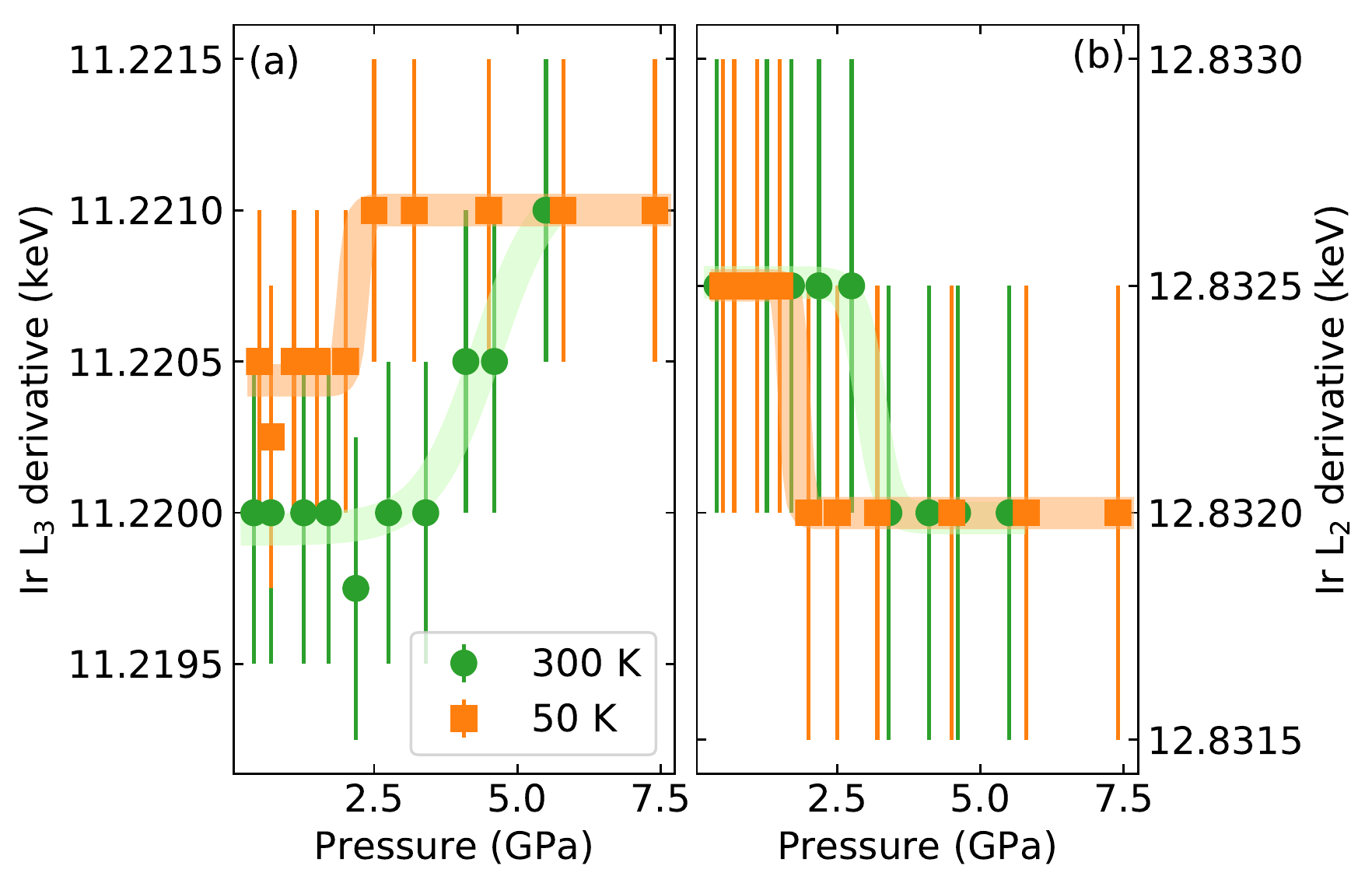}
	\caption{(Color online) Maximum value of the first derivative of the XANES spectra as a function of pressure at Ir (a) $L_3$ and (b) $L_2$ edges for both 300 K and 50 K. } 
	\label{AP5}
\end{figure}

Here, we present Ir $L_{2,3}$ XANES spectra and its first derivative as a function of pressure at $T=300$ K (Fig.~\ref{AP4}). The leading absorption edge shifts in opposite directions at $L_3$ and $L_2$ edges upon compression, indicating absence of a change in Ir valence (the energy at which the first derivative peaks is shown in Fig.~\ref{AP5}). That the Ir $5d$ occupation remains largely unchanged is also confirmed by the observation of a small change (reduction of 5\%) in the number of $5d$ holes at this temperature (see inset of Fig.~\ref{Figure6}(c)).


\section{Density Functional Theory calculations of the candidate crystal structures}
\label{DFT}

\begin{figure}
	\centering
	\includegraphics[width=\linewidth]{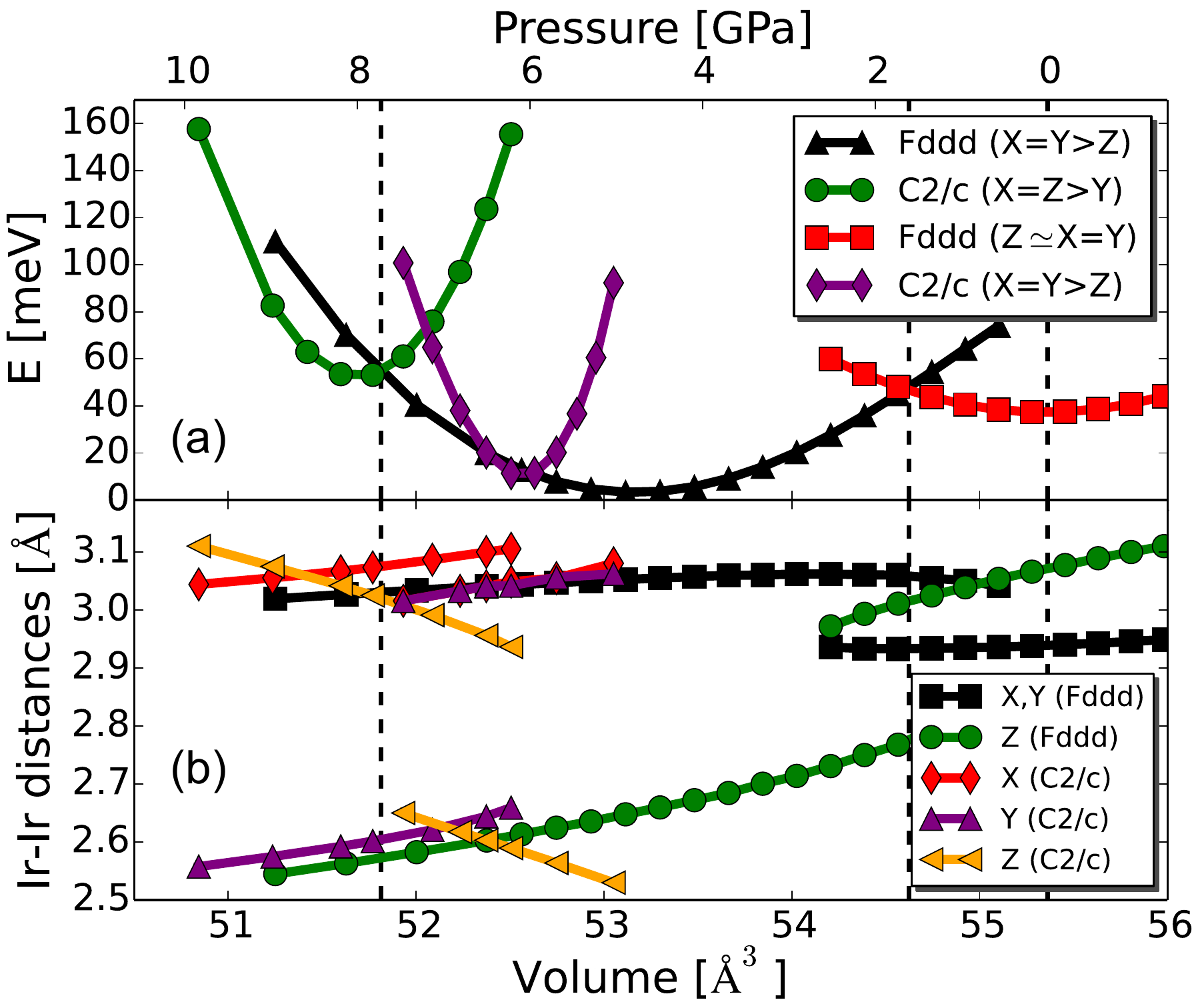}
	\caption{(Color online) (a) Total energy computed for different crystal structures in {\Li} as a function of volume and pressure. (b) Calculated Ir-Ir bond distances as a function of pressure and volume for each crystal structure, which reflects the changes of bond lengths through the phase transition boundaries. }
	\label{AP3}
\end{figure}

Density functional theory (DFT) calculations were used to study the energetics of candidate crystal structures for {\Li} under compression. We performed ionic relaxation and total energy calculations using the Vienna \emph{ab-initio} simulation package (VASP)~\cite{Kresse99, Kresse96}. The Perdew-Burke-Ernzerhof functional~\cite{Perdew96} was employed to treat the exchange-correlation and an energy cutoff of 600 eV was used. $k$-point meshes of $8\times6\times3$ and $8\times6\times6$ were used for the $Fddd$ and $C2/c$ structures studied here, respectively. Convergence of relaxed structures was achieved when forces for all atoms are smaller than 0.01 eV/{\AA}. While the inclusion of both correlation and spin-orbit (SO) effects can be important for precise structural determination and total energy calculations, a previous study on a similar honeycomb-lattice iridate shows that plain DFT is sufficient to capture lattice parameters and formation energies at the qualitative level~\cite{Manni2014}. Therefore, SO interaction and Hubbard $U$ correction were not included here.

The energetics of different crystal structures in compressed {\Li} were computed as function of unit cell volume (pressure). For each structure, the ion relaxation was performed by relaxing all internal atomic positions while the lattice parameters and the unit-cell shape were fixed. The total energy for each structure was computed using the relaxed structure. Since experimental structures are available only for selected crystal volumes (pressures), we obtain continuous sets of crystal volumes for each space group by interpolating lattice vectors from two different experimental structures. For the $Fddd$ structure, we interpolated from one structure measured at 0.5 GPa (single phase) and one at 2.5 GPa in a mixed phase. Since the $C2/c$ phase is stable only above 2 GPa in experiment, we interpolated the $C2/c$ structure using the 2.5 GPa and 3.5 GPa experimental structures.  

Figure~\ref{AP3}(a, b) show the total energy and Ir-Ir bond lengths computed for different crystal structures in {\Li} as a function of volume (pressure). Our calculations show that as the volume decreases a structural phase transition takes place between a $Fddd$ structure with no dimerization ($Z\sim X=Y$) (red curve) and a dimerized $Fddd$ structure with a shorter $Z$ bond ($X=Y>Z$), consistent with experiment (black curve). Further application of pressure produces the transition from $Fddd$ to $C2/c$ structure with a shorter $Y$ bond (green curve), also consistent with experiment. The black dashed lines in Fig.~\ref{AP3} show the theoretical phase boundaries. The ambient pressure phase in our DFT calculations is determined as the equilibrium volume of the $Fddd$ structure ($Z\sim X=Y$). The experimental bulk moduli for each phase (inverse of the compressibility) were used to convert the volume scale into a pressure scale. For example, the target pressure $P$ is computed as $P_0+B_0(V-V_0)/V_0$, where $P_0$ and $V_0$ are the initial pressure and volume of each phase boundary, $V$ is the target volume, and $B_0$ is the bulk modulus of each phase. 

While our calculations are qualitatively consistent with several experimental features, some differences are also noted compared to experiment. Although the $Fddd$ structure without dimerization becomes stable at larger volume, the total energy of that phase is not the global minimum. The lowest energy phase in our calculations is the dimerized $Fddd$ structure with a short $Z$ bond, only observed experimentally under compression. This discrepancy could originate in the neglect of SO interactions in our calculations since previous work has shown that a reduction in SO interaction can promote dimerization~\cite{Hermann2018}. Interestingly, our calculations find a metastable $C2/c$ phase with a short $Z$ bond, which is not observed experimentally. Nevertheless, this metastable phase is energetically very close to the dimerized $Fddd$ phase implying that competition between dimerized phases with different short bond selection may be at play. Moreover, the theoretical ambient pressure $Fddd$ phase shows a significant difference in $X$, $Y$ and $Z$ bond lengths while experiment shows almost equal Ir-Ir $X$, $Y$, $Z$ bond lengths.

\bibliography{references}

\end{document}